\documentclass[]{interact}

\usepackage{amsthm}
\usepackage[colorlinks=True,linkcolor=blue,citecolor=blue,urlcolor=blue,filecolor=blue,backref=page]{hyperref}
\usepackage{url}
\usepackage[utf8]{inputenc}
\usepackage{graphicx}
\usepackage{amsmath}
\usepackage{color}
\usepackage{multirow}
\usepackage[shortlabels]{enumitem}
\usepackage{xcolor}
\usepackage{amsfonts}

\usepackage{epstopdf}% To incorporate .eps illustrations using PDFLaTeX, etc.
\usepackage[caption=false]{subfig}% Support for small, `sub' figures and tables

\usepackage[numbers,sort&compress]{natbib}% Citation support using natbib.sty
\bibpunct[, ]{[}{]}{,}{n}{,}{,}% Citation support using natbib.sty
\renewcommand\bibfont{\fontsize{10}{12}\selectfont}% Bibliography support using natbib.sty

\theoremstyle{plain}% Theorem-like structures provided by amsthm.sty

\theoremstyle{definition}

\theoremstyle{remark}

\begin{document}

%\articletype{ARTICLE TEMPLATE}% Specify the article type or omit as appropriate

\title{Outcome-guided Bayesian Clustering for Disease Subtype Discovery Using High-dimensional Transcriptomic Data}

\author{
\name{Lingsong Meng and Zhiguang Huo\textsuperscript{*} \thanks{CONTACT Zhiguang Huo Email: zhuo@ufl.edu}}
\affil{\textsuperscript{*} Department of Biostatistics, University of Florida, Gainesville, Florida, U.S.A.}
}

\maketitle

\begin{abstract}

Due to the tremendous heterogeneity of disease manifestations, many complex diseases that were once thought to be single diseases are now considered to have disease subtypes.
Disease subtyping analysis, that is the identification of subgroups of patients with similar characteristics, is the first step to accomplish precision medicine.
With the advancement of high-throughput technologies, omics data offers unprecedented opportunity to reveal disease subtypes.
As a result, unsupervised clustering analysis has been widely used for this purpose.
Though promising, the subtypes obtained from traditional quantitative approaches may not always be clinically meaningful (i.e., correlate with clinical outcomes). 
On the other hand, the collection of rich clinical data in modern epidemiology studies has the great potential to facilitate the disease subtyping process via omics data and to discovery clinically meaningful disease subtypes.
Thus, we developed an outcome-guided Bayesian clustering (GuidedBayesianClustering) method to fully integrate the clinical data and the high-dimensional omics data.
A Gaussian mixed model framework was applied to perform sample clustering; a spike-and-slab prior was utilized to perform gene selection; a mixture model prior was employed to incorporate the guidance from a clinical outcome variable; and a decision framework was adopted to infer the false discovery rate of the selected genes.
We deployed conjugate priors to facilitate efficient Gibbs sampling.
Our proposed full Bayesian method is capable of simultaneously  (i) obtaining sample clustering (disease subtype discovery); (ii) performing feature selection (select genes related to the disease subtype); and (iii) utilizing clinical outcome variable to guide the disease subtype discovery.
The superior performance of the GuidedBayesianClustering was demonstrated through simulations and applications of breast cancer expression data.
An R package has been made publicly available on GitHub to improve the applicability of our method.
\end{abstract}

\begin{keywords}
Outcome-guided clustering; Bayesian method; Gaussian mixed model; Gibbs sampling
\end{keywords}

\section{Introduction}

Many complex diseases are difficult to treat because of the large amount of variabilities among the affected patients,
and the personalized medicine is a promising approach because of its potential to deliver the most responsive and effective therapy \citep{vogenberg2010personalized}. 
One of the most challenging and daunting tasks for developing personalized medicine is to perform disease subtyping -- identifying subgroups of patients with similar pathological conditions.
With the rapid advancement of high-throughput technology,
disease subtyping via molecular data (e.g., gene expression data) is becoming increasingly popular,
which has been applied to many diseases including 
lymphoma \citep{rosenwald2002use},
glioblastoma \citep{parsons2008integrated, verhaak2010integrated},
breast cancer \citep{lehmann2011identification, parker2009supervised},
colorectal cancer \citep{sadanandam2013colorectal},
ovarian cancer \citep{tothill2008novel},
Parkinson's disease \citep{williams2017parkinson} and
Alzheimer's disease \citep{bredesen2015metabolic}.

Taking breast cancer as an example, Parker et al.
\cite{parker2009supervised} developed 50 gene signatures (a.k.a PAM50) that classified breast cancer into five molecular subtypes, 
including Luminal A, Luminal B, Her2-enriched, Basal-like and Normal-like. 
These subtypes had shown distinct disease mechanisms, treatment responses and, survival outcomes \citep{van2002gene, coates2015tailoring}.
For example, 
the Luminal A subtype has the best prognosis, 
the HER2-enriched subtype can be treated by Herceptin,
and the Basal-like subtype is considered to have the worst survival.
The clinical value of these breast cancer molecular subtypes were further appreciated by clinical trial studies \citep{von2012definition, prat2015clinical}.

Unsupervised clustering methods, which aim to partition a dataset into several distinct subgroups, are effective ways to perform disease subtyping.
In the literature, several classical clustering methods have been employed for this purpose,
including  hierarchical clustering \citep{eisen1998cluster}, 
$K$-means \citep{dudoit2002prediction},
mixture model-based approaches 
\citep{mclachlan2002mixture}.
These classical clustering methods were particularly successful when the data is in low dimension (i.e., large number of samples and small number of genes).
Morden transcriptomic studies usually have tens of thousands of genes, 
and it is generally assumed that only a small subset of genes are related to the disease subtypes.
To accommodate this issue, 
sparse clustering algorithms were proposed to simultaneously select the intrinsic genes and perform sample clustering.
Along this direction, Witten and Tibshirani \cite{witten2010framework} proposed a sparse Kmeans algorithm.
In their paper, instead of assuming equal contribution of each gene feature that was used in the classical Kmeans, they designed a weighted Kmeans and imposed $l_1$/$l_2$ norm penalties on the gene weights. 
In their algorithm, the penalty would result in zero weights for many non-informative genes, and genes with non-zero weights were treated as selected genes.
Similarly, Pan and Shen \cite{pan2007penalized} and Xie et al. \cite{xie2008penalized} proposed to impose a weight penalization on the Gaussian mixture models. 
Bouveyron and Brunet-Saumard \cite{bouveyron2014model} provided a review for high-dimensional model-based clustering.

\begin{figure}[h!]
\centering
\includegraphics[width=1\textwidth]{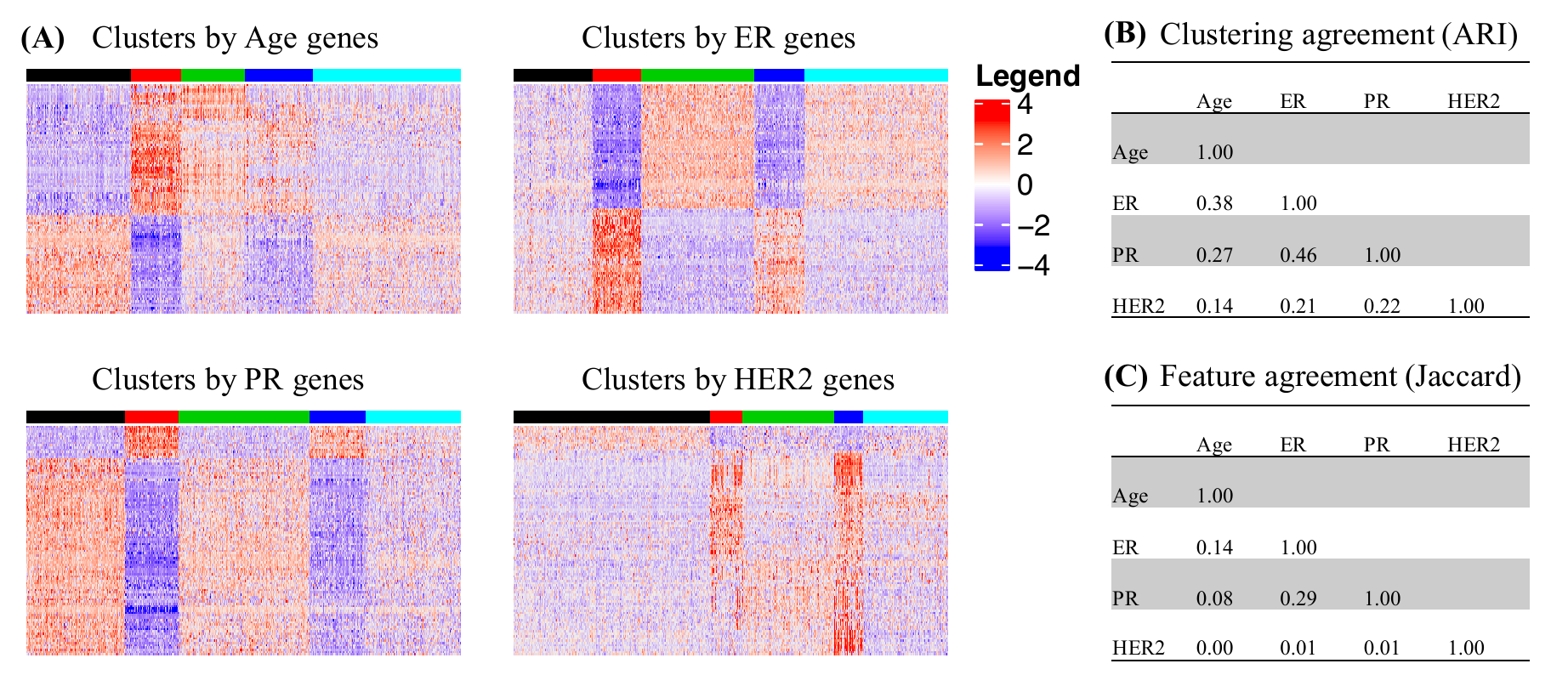}
\caption{Illustrating of multi-facet clusters defined by different genes. In (A), top 100 age-related, ER-related, PR-related, or HER2-related genes were used to perform clustering analysis, respectively. In the heatmap, each row represents a gene, and each column represents a sample. Within a heatmap, the samples under the same color bar represents the samples from the same cluster. (B) shows the pairwise cluster agreement among clustering results guided by different clinical variables. (C) shows the pairwise gene selection agreement among clustering results guided by different clinical variables.}
\label{fig:multiFacet}
\end{figure}

While these methods were successful in obtaining results for both clustering and gene selection, there are still limitations.
It is well acknowledged that clustering algorithms are sensitive to initializations and can be trapped in local optimum solutions.
Such local optimum problems can be further amplified in the case of high-dimensional data.
For high-dimensional data, people have noticed the existence of multi-facet clusters \citep{gaynor2017identification, nowak2008complementary},
where multiple configurations of sample clusters defined by separated gene sets may co-exist in the same dataset. 
These multiple configurations could be driven by genes associated with age, sex, and other confounding variables or pathological processes, rather than the intrinsic genes (i.e., genes related to the underlying disease).
We utilized the METABRIC data -- a breast cancer gene expression profile to illustrate the concept of the multi-facet clusters. 
This METABRIC data was also used in the later real data application (See Section~\ref{sec:metabric} for detailed description about this dataset).
Since age, estrogen receptor (ER), human epidermal growth factor receptor 2 (Her2), and  progesterone receptor (PR) were hallmarks of the Breast cancer,
we first pre-selected the top 100 significant age-related, ER-related, HER2-related, or PR-related genes via univariate regression.
Then, for each set of these pre-selected genes (e.g., top 100 ER-related genes), 
we extracted these genes from the high-dimensional gene expression profile as features, and performed sample clustering using the classical Kmeans.
This analysis was performed for each of these 4 sets of pre-mentioned breast cancer related genes, respectively.
Figure~\ref{fig:multiFacet}A showed that the subtype patterns obtained by different sets of pre-selected genes were quite distinct.
Figure~\ref{fig:multiFacet}B assessed the clustering agreement via ARI (See Section~\ref{sec:result} for definitions). 
The pairwise ARI ranged from 0.14 $\sim$ 0.47, indicating poor to moderate clustering agreement among configurations from different gene sets. 
Figure~\ref{fig:multiFacet}C compared the gene selection agreement via Jaccard index (See Section~\ref{sec:result} for definitions). 
The pairwise Jaccard indexes ranged from 0.00 $\sim$ 0.29, indicating poor gene selection agreement among clustering configurations from different gene sets. 
Collectively, Figure~\ref{fig:multiFacet} demonstrated the existence of multi-facet clusters
(distinct clustering configurations driven by different gene sets).
Therefore, without specifying disease-related genes, 
a clustering algorithm is likely to identify a subtype configuration that optimizes its objective function. 
However, the resulting subtype may not be clinically meaningful, and the selected genes may not be biologically relevant. For example, a clustering algorithm may identify a subtype configuration related to age, race, or gender, but not related to the specific disease of interest.

In modern biomedical studies, 
comprehensive clinical data are routinely collected.
Some of these information could be quite relevant to the underlying disease, 
and thus, properly incorporating such prior knowledge could potentially facilitate the identification of disease-related clustering configuration.
In the literature, several clustering methods have been proposed to incorporate prior knowledge.
Basu et al. \cite{basu2004active} proposed a constrained clustering algorithm by forcing/forbidding two samples in a cluster according to prior knowledge.
Huo et al. \cite{huo2017integrative} proposed an overlapping group lasso penalty to incorporate prior biological pathway information for a clustering algorithm.
However, these algorithms still could not ensure the resulting subtypes to be related to the underlying disease.
Bair and Tibshirani \cite{bair2004semi} proposed a two-step semi-supervised clustering methods,
where they pre-selected a list of disease-related genes, and then performed the regular Kmeans.
Though this algorithm emphasized that the selected genes were related to the disease,
the algorithm could not  promise that these selected genes were capable of minimizing within cluster dispersions.
In other word, the resulting subtypes could render large within group variability because the selected genes could carry large variability, 
which would make it difficult to predict a future patient.
In addition, they adopted an arbitrary cutoff to define their pre-selected genes, which may require further justifications in real data applications.
To address these limitations, we recently proposed an outcome-guided clustering framework \citep{meng2022outcome} by extending the sparse Kmeans algorithm (namely GuidedSparseKmeans). By using the guidance of clinical outcome variables, the GuidedSparseKmeans will obtain both clinically meaningful sample clustering and clinically relevant genes. 
Though successful, such frequentist’s method suffers from the following limitations: it only provide a single hard-threshold solution and does not enables probabilistic assignment of clustering membership. In addition, there is a lack of decision framework to reflect the uncertainty (i.e., false discovery rate) in feature selection.  
To address these limitations, unsupervised clustering analysis via Bayesian approach 
allows flexible statistical inference by generating a posterior distribution over the entire partition space. 

In the literature, several Bayesian clustering algorithms have been proposed including the Bayesian mixture model \citep{medvedovic2004bayesian}, the Bayesian non-parametric clustering model \citep{qin2006clustering}, and the Bayesian hierarchical clustering method \citep{heller2005bayesian}. To accommodate the high dimensional nature of the modern transcriptomic data, the sparse Bayesian clustering algorithm has been proposed \cite{luo2019batch}, where feature selections were accomplished by imposing spike and slab priors \citep{ishwaran2005spike}.

In this paper, 
we propose a full Bayesian hierarchical model to identify subtypes in a high-dimensional data, 
which will simultaneously (i) obtain sample clustering (disease subtype discovery); (ii) perform gene selection (select genes related to the disease subtype); and (iii) incorporate the guidance of a clinical outcome variable.
Utilizing disease-related clinical outcome guidance will encourage the identification of disease-related subtypes from the many configurations (multi-facet clusters) defined by other confounding genes.
In our model, a Gaussian mixture model framework is applied to perform sample clustering; a spike and slab prior is used for gene selection; a mixture model prior is employed to incorporate the guidance of a clinical outcome variable; and a decision framework is established to infer the false discovery rate of the selected genes.
Our full Bayesian framework has the advantage of providing a probabilistic belief of feature selection as well as soft assignment of cluster labels instead of a hard-thresholding approach. 
The priors are designed to be fully conjugate to facilitate efficient Gibbs sampling.
Our approach utilizes non-informative priors as much as possible such that the process of subtype identification process is data driven.
We evaluated the performance of our method in simulations and real data applications, 
and demonstrated its superior performance in comparison with regular sparse Bayesian clustering approach (without guidance term).
An R package has been made publicly available on GitHub to improve the applicability of our method.

\section{Method}
\label{sec:method}

\subsection{Gaussian mixture model} 
\label{sec:GaussianMixtureModel}
Denote $X_{gi}$  as the gene expression level of gene $g$ ($1\le g \le G$) for the sample $i$ ($1 \le i \le n$), where $G$ is total number of genes and $n$ is total number of samples.
We assume the gene expression matrix is properly standardized such that for each gene g, $\textbf{x}_g=(X_{g1},…,X_{gn})^\top$ has mean 0 and standardization 1.  

Denote  $Z_i$ as a subtype indicator, with $Z_i=k$ indicating sample $i$ belongs to subtype $k (1 \leq k \leq K)$, where $K$ is total number of subtypes. 
The scalar form $Z_i = k$ is equivalent to the vector form $Z_i=(0,…,0,1,0,…,0)^\top$, where 1 appears at the $k^{th}$ position. 
We will use these two forms interchangeably when there is no ambiguity.

By assuming (i) the gene expression data comes from a Gaussian mixture model, and (ii) genes are independent with each other, we have 
$$X_{gi} \sim \mbox{N}(\mu_{gk}, \sigma_g^2)|Z_i=k $$
$$Z_i \sim \mbox{Mult}(1; \pi_1, ..., \pi_K), $$
where $\mu_{gk}$ is the mean expression level of gene $g$ in subtype $k$;
$\sigma_{g}^2$ is variance of the expression level of gene $g$.
$\mbox{Mult}$ denotes the multinomial distribution; 
$\pi_{k}$ is the proportion of subtype $k$ and $\sum_{k=1}^K \pi_{k} = 1$;

Under the Guassian mixture model, the complete likelihood function for the observed data $\mathbf{X} = \{X_{gi}\}_{i=1,...,n; g=1,...,G}$ and subtype indicator $\mathbf{Z} = \{Z_{i}\}_{i=1,...,n}$ is:
$$ L(\mathbf{\Theta}|\mathbf{X}, \mathbf{Z}) = \prod_{i=1}^n \prod_{k=1}^K \left[ \pi_k \prod_{g=1}^G \frac{1}{\sqrt{2\pi}\sigma_g} \exp \left\{ -\frac{(X_{gi}-\mu_{gk})^2}{2\sigma_g^2} \right\} \right] ^{I(Z_i=k)} $$
where $\mathbf{\Theta}$ presents all unknown parameters including
$\pi_k$, $\mu_{gk}$ and $\sigma_g^2$ ($1\le k \le K$, $1 \le g \le G$),
and $I(\cdot)$ is an indicator function with $I(\cdot)=1$ if the expression inside $()$ is true and 0 otherwise.

\subsection{Sparse Gaussian mixture model} 
\label{sec:SparseGaussianMixtureModel}

Biologically, it is acknowledged that only a small subset of intrinsic genes will contribute to the final subtyping result. 
Recall that each gene has been standardized (i.e., $\sum_i X_{gi} = 0$).
Since an intrinsic gene should well separate different clusters, its cluster centers should be away from 0; 
while a non-intrinsic gene could not well separate different clusters, thus all of its cluster centers should be close to 0. 
We denote $L_g$ as the gene selection indicator, with $L_g=1$ indicating gene $g$ is selected, and $L_g=0$ indicating gene $g$ is not selected.
We denote $p$ as the prior probability of $L_g=1$. To achieve gene selection, we assume a spike-and-slab \citep{ishwaran2005spike} prior for $\mu_{gk}$: 
$$\mu_{gk} \sim \mbox{N}(0, \tau_{\mu 1}^2 )|L_g=1 ;$$ 
$$\mu_{gk} \sim \mbox{N}(0,\tau_{\mu 0}^2)|L_g=0 ;$$ 
$$L_g \sim \mbox{Bernoulli}(p) ,$$
where $\tau_{\mu 1}^2$ is the variance of the intrinsic genes, and $\tau_{\mu 0}^2$ is the variance of the non-intrinsic genes.
If $\tau_{\mu 1}^2 > \sigma^2_0$ ($\sigma^2_0$ is some positive number) and $\tau_{\mu 0}^2 \rightarrow 0$, 
then $\mu_{gk}|L_g=1$ is likely to be non-zero and $\mu_{gk}|L_g=0$ is likely to be close to 0.
By imposing this spike-and-slab prior, genes with large separation ability are likely to be selected (i.e., $ L_g=1$).
We will discuss how to specify $\tau_{\mu 1}^2$ and $\tau_{\mu 0}^2$ in section \ref{sec:prior}.
Such modeling strategy has been previously described by \cite{luo2019batch}.

\subsection{Guided Bayesian Clustering}

\subsubsection{Motivation of using mixture model priors to incorporate clinical outcome guidance} 
\label{sec:motivation}

\begin{figure}[h!]
\centering
\includegraphics[width=0.6\textwidth]{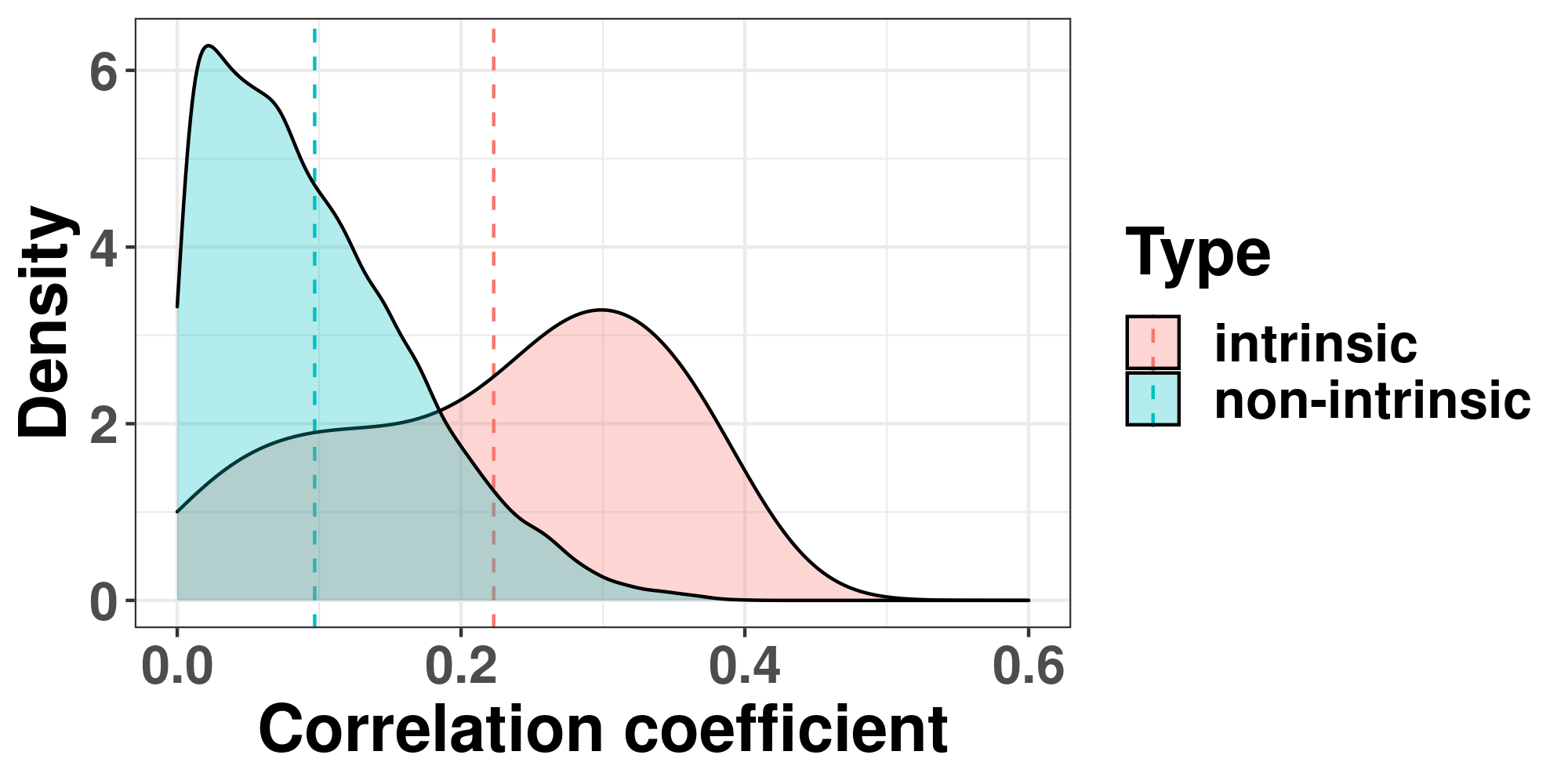}
\caption{Correlation coefficient distribution (absolute value) among intrinsic (selected) genes and non-intrinsic genes. Dashed lines represent the mean correlation.
The mean correlation for the intrinsic genes is 0.223, while the mean correlation for the non-intrinsic genes is 0.097. }
\label{fig:density}
\end{figure}

We hypothesize that a disease-related clinical variable has the potential to improve gene selection in a sparse clustering algorithm. 
To be specific, the intrinsic genes are more likely to be associated with the clinical variable than the non-intrinsic genes.
To examine this hypothesis, we first calculated the absolute values of  the correlation coefficient ($\boldsymbol{\rho} = (\rho_1, \ldots, \rho_G)^\top$) between a clinical variable (i.e., Nottingham prognostic index) and all gene features based on the METABRIC data (see Section~\ref{sec:metabric} for more details about this dataset). 
Here we defined intrinsic genes as the PAM50 genes \cite{parker2009supervised}, which was developed as a gold standard to classified breast cancer into five molecular subtypes.
As shown in Figure~\ref{fig:density}, the mean $\rho$ (in absolute value) in the selected genes group is 0.223, which is much higher than that of the non-selected genes group (0.097), with $p \leq 6.67 \times 10^{-9}$. 
This implies that we could potentially use these association strengths with respect to a clinical variable to facilitate gene selections.

To indicate the association strength between gene $g$ and a clinical outcome variable, we propose to introduce a gene-specific guide term $U_g = U(\textbf{x}_g, \textbf{y})$, 
where $\textbf{x}_g = (X_{g1}, \ldots, X_{gn})^\top \in \mathbb{R}^n$ is the expression levels of gene $g$, and 
$\textbf{y} = (y_1, \ldots, y_n)^\top \in \mathbb{R}^n$ is the vector of a clinical outcome variable. 
Under this definition, 
$\rho_g$ is a special case of  $ U_g $ when function $U$ represents the absolute value of the correlation coefficient. For the ease of modeling, we design and standardize the association strength $\textbf{u}=(U_1,…,U_G)^\top$ to range from (0,1). 
More discussion of the design on $\textbf{u}$ is available in Section \ref{sec:VariousOutcome}.

Motivated by Figure~\ref{fig:density}, we propose to model $\textbf{u}$ using a mixture model approach.
To be specific, we assume the distribution of $\textbf{u}$ is a mixture of two components. 
For the intrinsic genes,  the distribution of $\textbf{u}$ follows $F_1$, and for the non-intrinsic genes, the distribution of $\textbf{u}$ follows $F_0$.
Beta mixture model or truncated Gaussian mixture model are good candidates for this purpose.
To improve identifiability, we further impose a mean parameter shift between $F_0$ and $F_1$ to ensure $\mathbb{E}(F_1) > \mathbb{E}(F_0)$.
Though Beta mixture model is ideal for modeling the distribution of $\textbf{u}$, but it is lack of a closed-form solution for the posterior derivation.
For the convenience of Gibbs sampling, we adopt a truncated Gaussian mixture model with mean parameter 0 and truncation $[0, +\infty]$ for $F_0$ and mean parameter 1 and truncation $(-\infty, 1]$ for $F_1$ throughout our manuscript. 
This design will encourage that larger $U_g$ is more likely associated with $L_g=1$, and smaller $U_g$ is more likely associated with $L_g=0$.
This part can be further extended to other mixture models as needed.
The truncated Gaussian mixture model prior for $U_g$ is shown below:
$$U_g \sim \mbox{N}_{1-}(1, \tau_{U1}^2 )|L_g=1 ;$$
$$U_g \sim \mbox{N}_{0+}(0, \tau_{U0}^2 )|L_g=0 ,$$
where $\mbox{N}_{1-}$ is a truncated normal distribution with right-side truncation at 1; $\mbox{N}_{0+}$ is a truncated normal distribution with left-side truncation at 0; $\tau_{U1}^2$ is the variance parameter for the association strength of the intrinsic gene component (i.e., $L_g=1$) and $\tau_{U0}^2$ is the variance parameter for the association strength of the non-intrinsic gene component (i.e., $L_g=0$).
The selection of $\tau_{U1}^2$ and $\tau_{U0}^2$ will be discussed in Section~\ref{sec:prior}.
Such a mixture model will encourage the selection of genes that are highly associated with the clinical outcome variable.

\subsubsection{Full Bayesian model}

\begin{figure}[h!]
\centering
\includegraphics[scale=0.28]{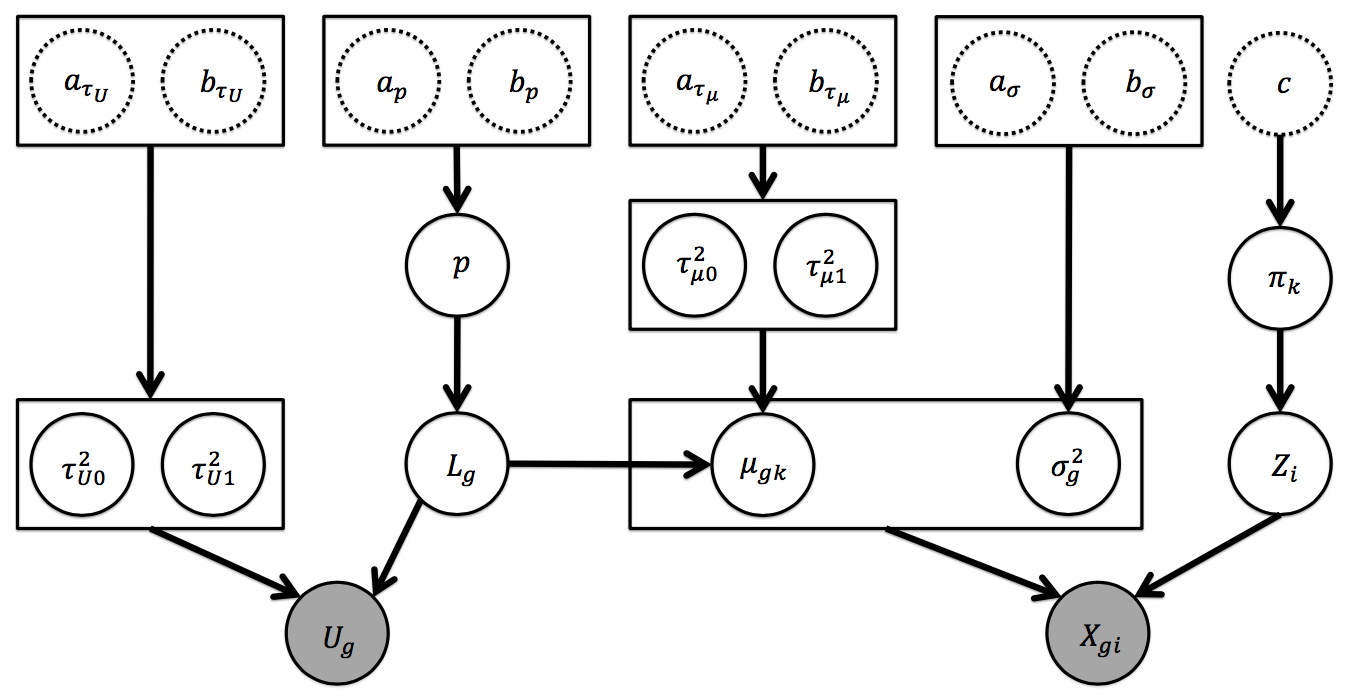}
\caption{Graphical representation of Bayesian latent hierarchical model.
Shaded nodes indicate observed variables. Dashed nodes indicate pre-fixed hyper parameters. Arrows show generative process. $g (1 \le g \le G)$ is the gene index, $i (1 \le i \le n)$ is the sample index, and $k (1 \le k \le K)$ is the subtype index.}
\label{fig:graphical}
\end{figure}

Figure~\ref{fig:graphical} shows the graphical model representation of the data generative process of our Bayesian latent hierarchical model. 
Our model is consisted of three major components. 
(i) The right component, including $Z_i$, $X_{gi}$, $\mu_{gk}$, and $\sigma_g^2$, represents the Gaussian mixture model introduced in Section~\ref{sec:GaussianMixtureModel}.
This component is responsible for inferring the clustering results (i.e., $Z_i$).  
(ii) The middle component, including $L_g$, $\mu_{gk}$, $\tau_{\mu0}^2$ and $\tau_{\mu1}^2$, represents the spike-and-slab prior introduced in Section~\ref{sec:SparseGaussianMixtureModel}.
This component is responsible for performing gene selection (i.e., select genes with large separation ability).
(iii) The left component, including $L_g$, $U_{g}$, $\tau_{U0}^2$ and $\tau_{U1}^2$, represents the mixture model prior introduced in Section~\ref{sec:motivation}.
This component is responsible for incorporating clinical outcome information to facilitate the disease subtyping process.
The parameters of interest $\mathbf{\Theta}$  include  
$\mu_{gk}$, $\sigma_g^2$, $L_g$, $Z_i$, $p$, $\pi_k$, $\tau_{\mu0}^2$, $\tau_{\mu1}^2$, $\tau_{U0}^2$, $\tau_{U1}^2$, where $1\le k \le K$, $1 \le g \le G$.
By inferring $\mathbf{\Theta}$ from this full Bayesian model, we will simultaneously (i) obtain sample clustering result; (ii) select genes with strong separation ability; and (iii) utilize a clinical outcome variable to enhance gene selection and sample clustering.

Based on this graphical model, the full posterior likelihood $L ( \mathbf{\Theta} | \bf{X}, \bf{u})$ is proportional to:
\label{FullPostLikelihood}
\begin{align*}
& 
\prod_{i=1}^n \prod_{k=1}^K \left[\pi_k \prod_{g=1}^G \frac{1}{\sqrt{2\pi}\sigma_g} \exp \left\{ -\frac{(X_{gi}-\mu_{gk})^2}{2\sigma_g^2} \right\} \right] ^{I(Z_i=k)} \\
& \cdot
\prod_{g=1}^G \left[\mbox{N}_{1-}(U_g; 1, \tau_{U1}^2) \cdot L_g +
\mbox{N}_{0+}(U_g; 0, \tau_{U0}^2) \cdot (1-L_g)\right] \\
& \cdot 
\prod_{g=1}^G \prod_{k=1}^K \left[\mbox{N}(\mu_{gk}; 0, \tau_{\mu1}^2) \cdot L_g +
\mbox{N}(\mu_{gk}; 0, \tau_{\mu0}^2) \cdot (1-L_g)\right] \\
& \cdot
\prod_{g=1}^G 
\left[\mbox{Inv}\Gamma(\sigma_g^2, a_{\sigma}, b_{\sigma}) \cdot
p^{L_g} (1-p)^{1-L_g} \right] \cdot 
\mbox{Beta}(p; a_p,b_p) \cdot
\mbox{Dir}(\boldsymbol{\pi}; \mathbf{c}) \\
& \cdot
\mbox{Inv}\Gamma(\tau_{\mu1}^2; a_{\tau_{\mu1}}, b_{\tau_{\mu1}}) \cdot
\mbox{Inv}\Gamma(\tau_{\mu0}^2; a_{\tau_{\mu0}}, b_{\tau_{\mu0}}) \cdot 
\mbox{Inv}\Gamma(\tau_{U1}^2; a_{\tau_{U1}}, b_{\tau_{U1}}) \cdot
\mbox{Inv}\Gamma(\tau_{U0}^2; a_{\tau_{U0}}, b_{\tau_{U0}}) ,
\end{align*}
where some of the priors and hyper parameters were introduced in the following sections.

\subsubsection{Prior specification}
\label{sec:prior}

We apply independent conjugate priors to each component in $\mathbf{\Theta}$ as follows: $\boldsymbol{\pi} = (\pi_1,...,\pi_K)^\top \sim \mbox{Dir}(\mathbf{c})$; 
$\sigma_g^2 \sim \mbox{Inv}\Gamma(a_{\sigma}, b_{\sigma})$;
$p \sim \mbox{Beta}(a_p, b_p)$;
$\tau_{\mu1}^2 \sim \mbox{Inv}\Gamma(a_{\tau_{\mu1}}, b_{\tau_{\mu1}})$;
$\tau_{\mu0}^2 \sim \mbox{Inv}\Gamma(a_{\tau_{\mu0}}, b_{\tau_{\mu0}})$;
$\tau_{U1}^2 \sim \mbox{Inv}\Gamma(a_{\tau_{U1}}, b_{\tau_{U1}})$;
$\tau_{U0}^2 \sim \mbox{Inv}\Gamma(a_{\tau_{U0}}, b_{\tau_{U0}})$.
Such design of conjugate priors will greatly facilitate efficient implementations of Gibbs sampling.
Here, $\mathbf{c} = (c,\ldots,c)^\top \in \mathbb{R}^K$, $a_{\sigma}$, $b_{\sigma}$, $a_p$, $b_p$, $a_{\tau_{\mu1}}$, $b_{\tau_{\mu1}}$, $a_{\tau_{\mu0}}$, $b_{\tau_{\mu0}}$, $a_{\tau_{U1}}$, $b_{\tau_{U1}}$, $a_{\tau_{U0}}$ and $b_{\tau_{U0}}$ are hyper parameters.

\subsubsection{Hyper parameter justification}
We propose to assign non-informative prior wherever possible. For instance, we will set $c=1; a_p=b_p=1; a_{\sigma}=b_{\sigma}=0.001; a_{\tau_{U0}}=b_{\tau_{U0}}=0.001; a_{\tau_{U1}}=b_{\tau_{U1}}=0.001$. 
In order to distinguish intrinsic genes and non-intrinsic genes (make them identifiable), we set $a_{\tau_{\mu}}$ and $b_{\tau_{\mu}}$ to be informative. 
To be specific, we will set $a_{\tau_{\mu0}}=2$, $b_{\tau_{\mu0}}=0.005$, which will result in a small prior mean for $\tau^2_{\mu0}$; $a_{\tau_{\mu1}}=4$, $b_{\tau_{\mu1}}=450$, which will result in large prior mean for $\tau^2_{\mu1}$. 
The justification for these informative hyper parameters will be done by sensitivity analysis (see Section~\ref{sec:sens}). 
We use the Bayesian information criterion (BIC) \citep{schwarz1978estimating} to choose the number of subtypes $K$.
The BIC formula for the GuidedBayesianClustering  \citep{schwarz1978estimating} is 
$$ -2 \left[ \sum_{i=1}^n \log \left( \sum_{k=1}^K \hat{\pi}_k \prod_{g=1}^G \mbox{N}(X_{gi}; \hat{\mu}_{gk}, \hat{\sigma}_g^2 ) \right)  \right] + KG \log G , $$
where the first term is negative two times the log likelihood; the second term is the product of the parameter number and the logarithm of number of genes; $\hat{\pi}_{k}, \hat{\mu}_{gk}$ and $\hat{\sigma}_g^2$ are the posterior mean estimates.
The number of subtypes $K$ is chosen with the minimum BIC.

\subsubsection{Posterior Calculation}
\label{sec:post}

We develop a Gibbs sampler algorithm to draw samples \citep{geman1984stochastic, robert2013monte} for $\mathbf{\Theta}$. 
To be specific, in each Gibbs sampling iteration, we update one parameter in $\mathbf{\Theta}$ while conditioning on all other parameters at their most recently updated value.
The order for updating the parameter is fixed as $p$, $\tau_{\mu0}^2$, $\tau_{\mu1}^2$, $\tau_{U0}^2$, $\tau_{U1}^2$, $L_g$, $\boldsymbol{\pi}$, $Z_i$, $\mu_{gk}$, $\sigma_g^2$,
where $1\le k \le K$ and $1 \le g \le G$.

1.  Update the proportion of intrinsic genes $p$ from
$$\mbox{Beta} \left(a_p + \sum_{g=1}^{G} L_g , b_p + G - L_g \right) .$$

2. Sample the variance of the spike component of the spike-and-slab prior $\tau_{\mu0}^2$ from
$$ \mbox{Inv}\Gamma \left(a_{\tau_{\mu0}} + \frac{K}{2} \sum_{g=1}^G I( L_g=0), b_{\tau_{\mu0}} + \frac{1}{2} \sum_{\{(g,k):L_g=0\}} \mu_{gk}^2 \right) .$$

3. Sample the variance of the slab component of the spike-and-slab prior $\tau_{\mu1}^2$ from
$$ \mbox{Inv}\Gamma \left(a_{\tau_{\mu1}} + \frac{K}{2} \sum_{g=1}^G I( L_g=1), b_{\tau_{\mu1}} + \frac{1}{2} \sum_{\{(g,k):L_g=1\}} \mu_{gk}^2 \right) .$$

4. Sample the variance of the guidance of non-intrinsic genes  $\tau_{U0}^2$ from
$$ \mbox{Inv}\Gamma \left(a_{\tau_{U0}} + \frac{1}{2} \sum_{g=1}^G I( L_g=0), b_{\tau_{U0}} + \frac{1}{2} \sum_{\{g:L_g=0\}} U_{g}^2 \right) .$$

5. Sample the variance of the guidance of intrinsic genes  $\tau_{U1}^2$ from
$$ \mbox{Inv}\Gamma \left(a_{\tau_{U1}} + \frac{1}{2} \sum_{g=1}^G I( L_g=1), b_{\tau_{U1}} + \frac{1}{2} \sum_{\{g:L_g=1\}} (U_{g}-1)^2 \right) .$$

6. Update gene selection indicator $L_g$ from the Bernoulli distribution
$$ \mbox{Bern} \left( \frac{p \cdot \prod_{k=1}^{K}  \mbox{N}(\mu_{gk}; 0, \tau_{\mu1}^2)  \cdot \mbox{N}(U_{g}; 1, \tau_{U1}^2)} 
{p \cdot \prod_{k=1}^{K} \mbox{N}(\mu_{gk}; 0, \tau_{\mu1}^2)  \cdot \mbox{N}(U_{g}; 1, \tau_{U1}^2) + (1-p) \cdot \prod_{k=1}^{K} \mbox{N}(\mu_{gk}; 0, \tau_{\mu0}^2) \cdot  \mbox{N}(U_{g}; 0, \tau_{U0}^2)} \right) .$$

7. Sample subtype proportion $\boldsymbol{\pi}$ from the Dirichlet distribution
$$ \mbox{Dir} \left(c + \sum_{i=1}^{n} I(Z_i=1), ..., c + \sum_{i=1}^{n} I(Z_i=K)\right) .$$

8. For each sample $i$, update its subtype indicator $Z_i$ based on multinomial distribution
$$\mbox{Mult}(1; q_1, \ldots, q_K) ,$$
where $q_k^* = \pi_k \exp \left\{-\sum_{g=1}^G \frac{(X_{gi} - \mu_{gk})^2} {2\sigma_g^2} \right\} $ and $ q_k = \frac{q_k^*}{\sum_k q_k^* } .$

9. For each gene $g$ and each subtype $k$, sample the mean of gene expression $\mu_{gk}$ from

$$ \mbox{N} \left( \frac{\tau_{\mu L_g}^2 \sum_{i \in \{Z_i=k\}} X_{gi} \sigma_g^2} { \tau_{\mu L_g}^2 \cdot \sum_{i=1}^n I(Z_i=k) + \sigma_g^2},  
\frac{\tau_{\mu L_g}^2 \sigma_g^2} { \tau_{\mu L_g}^2 \cdot \sum_{i=1}^n I(Z_i=k) + \sigma_g^2} \right) ,$$
where $\tau_{\mu L_g} = I(L_g=1) \tau_{\mu 1} + I(L_g=0) \tau_{\mu 0} .$

10. For each gene, sample the variance of gene expression $\sigma_g^2$ from
$$ \mbox{Inv}\Gamma \left(a_{\sigma} + \frac{n}{2}, b_{\sigma} + \frac{1}{2} \sum_{i=1}^{n} (X_{gi} - \mu_{g Z_i})^2 \right) ,$$
where $\mu_{g Z_i} = \mu_{g1}$ if $Z_i = 1$ and $\mu_{g Z_i} = \mu_{g0}$ if $Z_i = 0$.

\subsubsection{Decision making}

Denote $N_T$ as the total number of iterations from the Gibbs sampling; $N_B$ as the number of burn-in samples.
The  burn-in samples are discarded from the Bayesian inference because these initial samples may not necessarily converge to the stationary distribution of the full posterior likelihood (Equation \ref{FullPostLikelihood}).
Throughout this manuscript, we set $N_T = 3000$ and $N_B = 1500$, unless otherwise specified.
After the Gibbs sampling, a total of $N = N_T – N_B$ posterior samples are used for further Bayesian inference. 

To infer if a gene is an intrinsic gene (i.e., genes that contribute to separate the subtypes), we first denote $\Omega_I$ as the collection of intrinsic genes (i.e., $\Omega_I = \{g: 1\le g \le G$; gene $g$ is an intrinsic gene$\}$), and $\Omega_{\overline{I}}$ as the collection of non-intrinsic genes (i.e., $\Omega_{\overline{I}} = \{g: 1\le g \le G; g \notin \Omega_I \}$). 
We denote $P_g = \mbox{Pr}(g \in \Omega_{\overline{I}} | L_g = 1) = 1 - \mbox{Pr} (g \in \Omega_I | L_g = 1) $, which is also referred as the local false discovery rate \citep{efron2002empirical}.
Given a threshold $\eta$ ($0 < \eta < 1$.), when claiming gene $g$ as an intrinsic gene if $P_g \le \eta$, the expected number of false discoveries is $\sum_g P_g I(P_g \le \eta)$.
According to Newton et al. \cite{newton2004detecting}, the resulting expected false discovery rate for genes with  $P_g \le \eta$, $1 \le g \le G$ is
$$\mbox{FDR}(\eta) = 
\frac{\sum_{g=1}^G P_g \cdot I(P_g \leq \eta)} {\sum_{g=1}^G I(P_g \leq \eta)} .$$
In practice, we will estimate $ P_g $ as $1 - \frac{1}{N} \sum_{t=N_B+1}^{N_T} L_g^{[t]}$.

We infer the subtype for each subject $i$ from the posterior samples of $Z_i$.
A probabilistic assignment of sample $i$ to cluster $k$ can be calculated as 
$$l_i^{(k)} = \frac{1}{N} \sum_{t= N_B+1}^{N_T} I(Z_i^{[t]} = k).$$
In our paper, we used the maximum a posteriori (MAP) estimation to decide the cluster assignment for sample $i$ (i.e., $\arg \max_k l_i^{(k)}  $).
To solve potential label switching problems, we adopted pivotal reordering algorithm by Marin et al. \citep{marin2005bayesian, marin2007bayesian}.

\subsubsection{Extension to other types of clinical outcome variables} 
\label{sec:VariousOutcome}
In Section~\ref{sec:motivation}, 
a gene specific guidance term $U_g$ ($0 \le U_g \le 1$) was utilized to measure the strength of association between gene $g$ and a clinical outcome variable.
If the clinical variable is continuous, we can compute $U_g$ as the absolute value of the Pearson correlation ($\rho$) between gene $g$ and the clinical outcome variable, 
or the coefficient of determination $R^2$ ($R^2$ is the same as the square of $\rho$) from a univariate linear regression model, where the dependent variable is the clinical variable and the independent variable is the expression level of gene $g$.
In general, the clinical variable can be of any data type, including continuous, binary, ordinal, count, survival, etc.
We extend the linear regression model to a generalized univariate regression model $f_g$ to accommodate clinical outcome variables with various types.
For example, 
generalized linear models can be used for binary, ordinal, and count data;
Cox models can be used for survival data.
Similar to the coefficient of determination $R^2$,
Cox and Snell \cite{cox1989analysis} proposed the pseudo R-squared for a generic univariate regression $f_g$:
$$R_g^2 = 1 - \Bigl[ \frac{L(f_0)}{L(f_g)} \Bigr] ^{2/n} $$
where $L(f_0)$ is the likelihood of null model; $L(f_g)$ is the likelihood of the model $f_g$; and $n$ is the number of subjects. 
To ensure this term has the scale of [0,1], we further proposed an adjusted pseudo R-squared:
$$U_g = \frac{R_g^2 - \min(R_g^2)}{\max(R_g^2) - \min(R_g^2)} $$

\section{Result}
\label{sec:result}

In this section, we first evaluated the performance of the GuidedBayesianClustering using simulation datasets, and compared with the regular sparse Bayesian clustering method (the BayesianClustering, see Section~\ref{sec:method}).
Note that we don’t compare with any frequentist clustering methods because they cannot enable statistical inference (i.e., obtain the false discovery rate of the selected genes). 
Further, we applied these methods in a gene expression profile of breast cancer to illustrate the superior performance of our proposed method. 
We benchmarked the performance in terms of both clustering performance and gene selection performance. 
For the clustering performance, we used adjusted Rand index \citep{hubert1985comparing} (ARI). 
ARI characterizes the consistency between a clustering assignment result and the underlying true clustering assignment,
which ranges from -1 (indicating poor agreement) to 1 (indicating perfect agreement).
For gene selection performance,
we  used the Jaccard index \citep{jaccard1901etude} to measure the similarity between the selected genes and the intrinsic genes.
The Jaccard index was defined as the ratio of the number of intersecting genes occurring in two genomes to the number of genes occurring in at least one genome.
The range for a Jaccard index is from 0 (indicating no overlap) to 1 (indicating fully overlap).

\subsection{Simulation} 
\label{sec:simu}

\subsubsection{Simulation setting}

A gene expression study with $K=3$ subtypes was simulated to evaluate the performance of the GuidedBayesianClustering and compare it with the BayesianClustering (i.e., the sparse Gaussian mixture model, Section~\ref{sec:SparseGaussianMixtureModel}).
To mimic the multifaceted clustering configurations defined by different gene sets, we simulated intrinsic genes that define disease-related subtype clusters, confounding influence genes that define other clustering configurations (not related to disease), and noise genes (i.e., housekeeping genes).
We modeled correlated gene structures for intrinsic genes and confounding impacted genes to best capture the complex structure of genomic data.
Additionally, a continuous outcome variable related to the intrinsic genes was generated as the clinical guidance. 
Below is the detailed simulation data generation process, which was also similarly described elsewhere \citep{huo2016meta, huo2017integrative, meng2022outcome}.

\begin{enumerate}[(a)]
	\item Intrinsic genes.
    \begin{enumerate}[1.]
	    \item Generate $K=3$ disease-related subtypes. Generate $N_k \sim \mbox{POI}(100)$ patients for each subtype $k (1 \le k \le K)$, where $\mbox{POI}$ indicates a Poisson distribution. In the simulation, the total number of patients is $N=\sum_k N_k$.
	    
	    \item Generate $M=20$ gene modules. Generate $n_m \sim \mbox{POI}(20)$, where $n_m$ indicates the number of gene features of module $m$ ($1 \le m \le M$). Repeat this procedure for all $M$ modules, which will result in 400 intrinsic genes on average. 
        
	    \item Denote the baseline level of subtype $k(1 \le k \le K)$ as $\theta_k$, and the template gene expression for subtype $k$ and module $m$ level as $\mu_{km}$. 
        Then, the baseline level is calculated as $\theta_k = 2 + 2k$, and the template gene expression is generated by $\mu_{km} = \alpha_m \theta_k + \mbox{N}(0, \sigma_0^2)$, where $\alpha_m \sim \mbox{UNIF} \bigl( (-2, -0.2) \cup (0.2, 2) \bigr)$ indicates the fold change for each module $m$; and $\sigma_0$ is fixed to be 1.
	    
	    \item  Impose biological variation $\sigma_1^2$ to the template gene expression $\mu_{km}$ such that $X_{kmi}^{'} \sim \mbox{N}(\mu_{km}, \sigma_1^2)$, where $k$ indicates the subtype index, $m(1 \le m \le M)$ indicates the module index, and $i(1 \le i \le N_k)$ indicates the patient index. We fix $\sigma_1$ to be 3 unless otherwise specified.
	    
	    \item Impose correlation structure for genes in subtype $k$ and module $m$. First, generate $\Sigma_{km}^{'} \sim \mbox{W}^{-1}(\phi, \nu)$, where $\mbox{W}^{-1}$ indicates the inverse Wishart distribution, $\phi = 0.5I_{n_m \times n_m} + 0.5J_{n_m \times n_m}$, $ \nu = 60$, $I_{n_m \times n_m}$ is an $n_m$ by $n_m$ identity matrix and $J_{n_m \times n_m}$ is an $n_m$ by $n_m$ matrix with all elements equal to 1. 
        The covariance matrix $\Sigma_{km}$ is computed via standardizing  $\Sigma_{km}^{'}$ such that all the diagonal elements are equal to 1.
        
	    \item Generate gene expression values for all genes in module $m$ as $(X_{1kmi}, ..., X_{n_mkmi})^\top \sim \mbox{MVN}(X_{kmi}^{'}, \Sigma_{km})$, where $1 \le k \le K$, $1 \le m \le M$ and $1 \le i \le N_k$. $\mbox{MVN}$ indicates the multivariate normal distribution.

    \end{enumerate}
    \item Phenotypic variables.
    \begin{enumerate}[1.]
	    \item Generate the continuous clinical outcome variable as $Y_{ki} \sim \mbox{N}(\theta_k, \sigma_2^2)$ for subject $i$ ($1 \le i \le N_k$) in subtype $k$ ($1 \le k \le K$). 
        We fix $\sigma_2$ to be 6 such that the pattern of the guidance term $U_g$ of the intrinsic genes in the simulation can be compared to that in the breast cancer example (See details in Section~\ref{sec:metabric}).
	\end{enumerate}
    \item Confounding impacted genes.
    \begin{enumerate}[1.]
	    \item Generate $V=4$ confounding variables. 
        Confounding variables could be age, sex, race, or other confounding factors that could define non-disease associated subtype clusters. These variables may complicate the process of discovering disease subtypes.
        We similarly generate $R=20$ modules for each confounding variable $v(1 \le v \le V)$, and sample number of genes $n_{r_v} \sim \mbox{POI}(20)$ for each module $r_v(1 \le r_v \le R)$. 
        There are 1,600 confounding impacted genes on average after repeating this procedure for all modules in all confounding variables.
        
        \item Randomly divide the $N$ samples into $K$ subclasses for each confounding variable $v$, representing the non-disease-related clusters defined by confounding impacted genes.
        
        \item Similar to Step a3, set the baseline gene expression of subclass $k(1 \le k \le K)$ for each confounding variable $\theta_{k} = 2+2k$.         
        Denote $\mu_{kr_c}$ as the template gene expression of subclass $k(1 \le k \le K)$ and module $r_c(1 \le r_c \le 20)$. 
        The template gene expression is calculated by $\mu_{kr_c} = \alpha_{r_c} \theta_k + \mbox{N}(0, \sigma_0^2)$, where $\alpha_{r_c}$ indicates the fold change for each module $r_c$ and $\alpha_{r_c} \sim \mbox{UNIF} \bigl( (-2, -0.2) \cup (0.2, 2) \bigr)$.  
        
        \item Impose the biological variation $\sigma_1^2$ to the template gene expression such that $X_{kr_ci}^{'} \sim \mbox{N}(\mu_{kr_c}, \sigma_1^2)$. 
        \item Following Step a5 and a6, we simulate gene correlation structure $\Sigma_{km}$ within modules of confounder impacted genes, and generate their gene expression by $(X_{1kr_ci}, ..., X_{n_m k r_c i})^\top \sim \mbox{MVN}(X_{kr_ci}^{'}, \Sigma_{km})$.  
	\end{enumerate}
    \item Noise genes.
    \begin{enumerate}[1.]
	    \item Generate additional 3,000 non-informative noise genes (i.e., housekeeping genes) denoted by $g$. 
        We generate template gene expression $\mu_g \sim \mbox{UNIF}(4, 8)$. 
        Then, we impose noise $\sigma_3 = 1$ to the template gene expression and generate $Y_{gi} \sim \mbox{N}(\mu_g, \sigma_3^2)$.
	\end{enumerate}
\end{enumerate}

\subsubsection{Simulation results}

\begin{figure}
\centering
\subfloat[Selection of K based on BIC]{%
\resizebox*{4.6cm}{!}{\label{fig:bic} \includegraphics{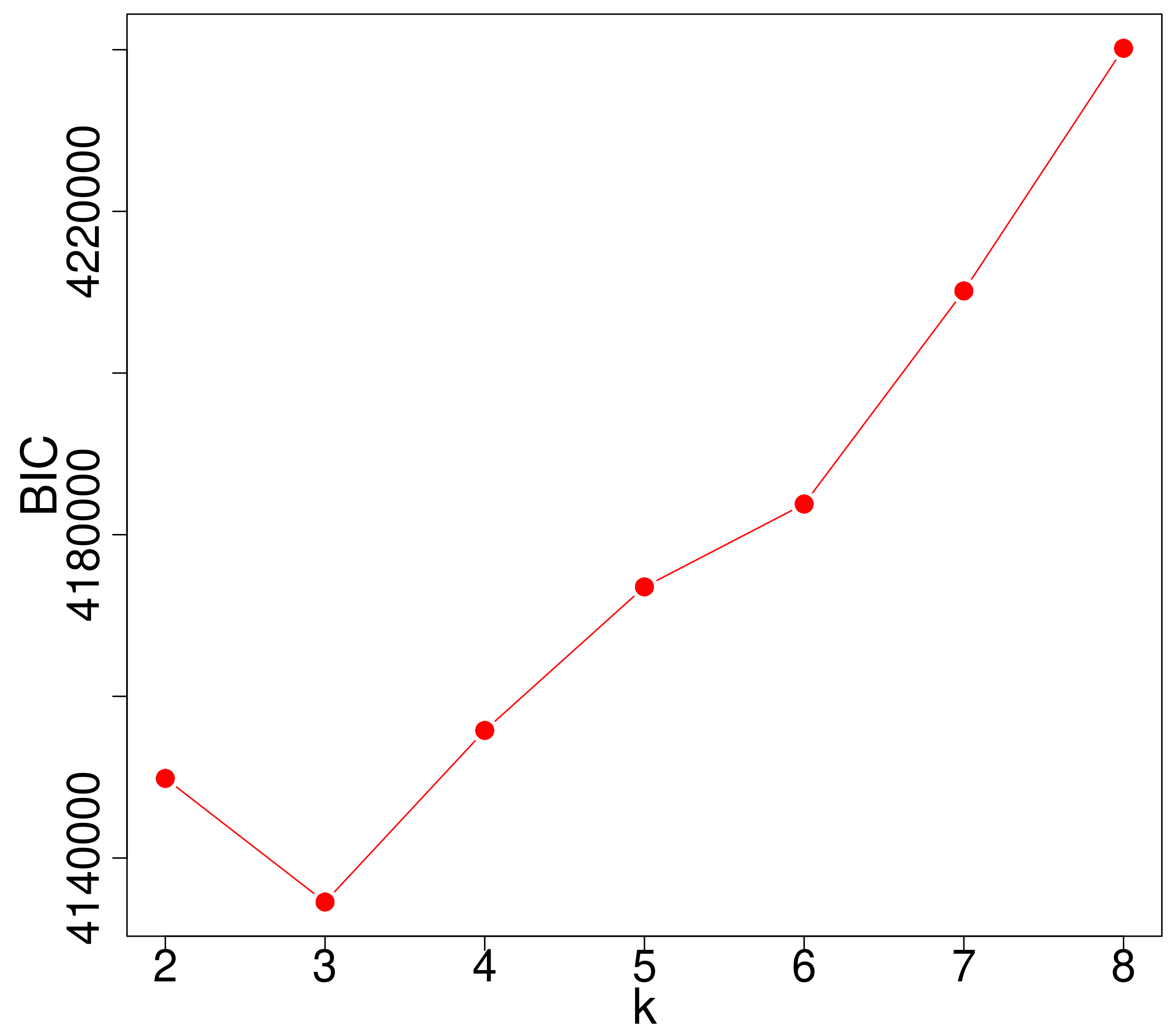}}}\hspace{5pt}
\subfloat[Clustering results]{%
\resizebox*{4.6cm}{!}{\label{fig:cluster} \includegraphics{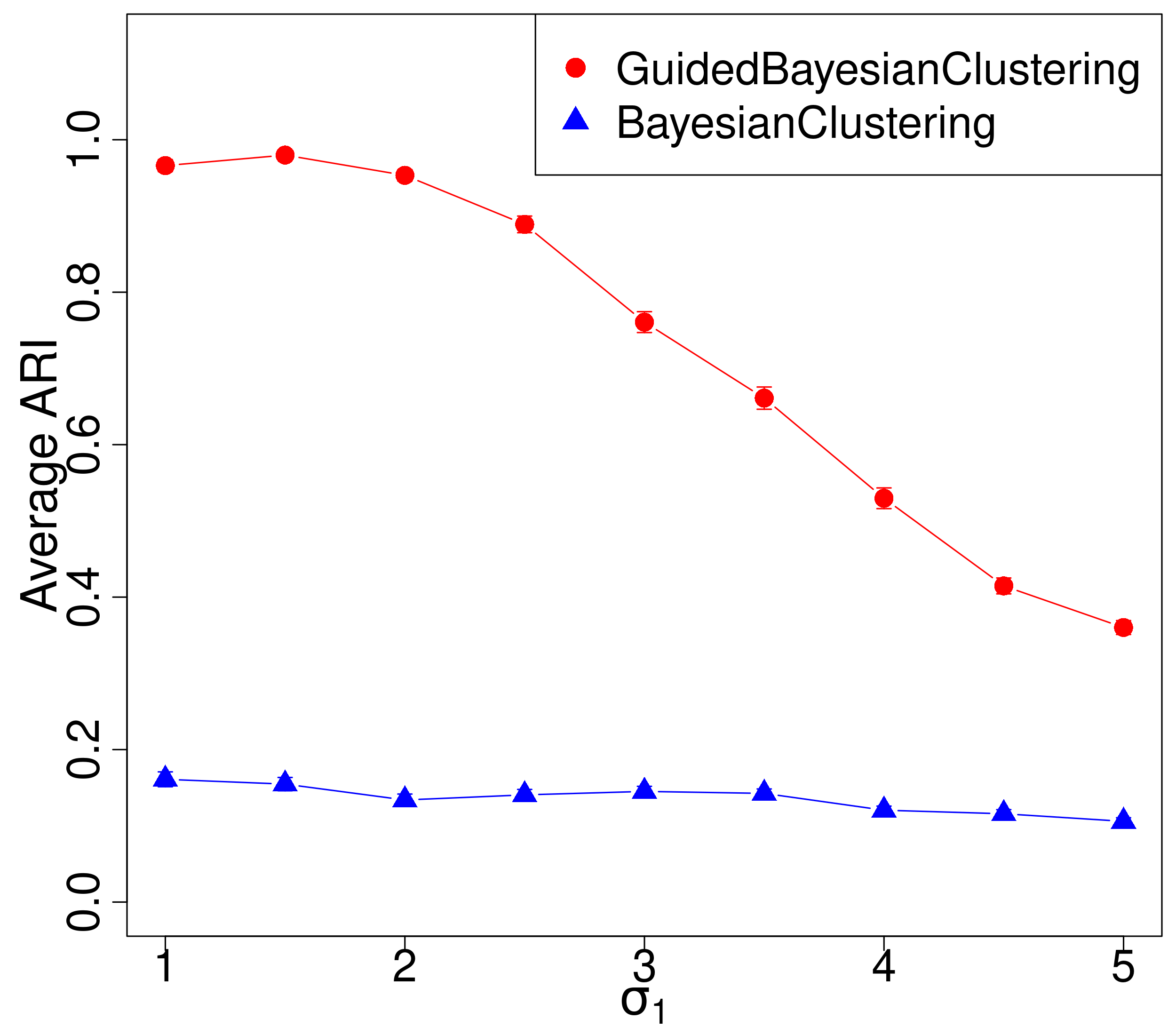}}}
\subfloat[Gene selection results]{%
\resizebox*{4.6cm}{!}{\label{fig:gene} \includegraphics{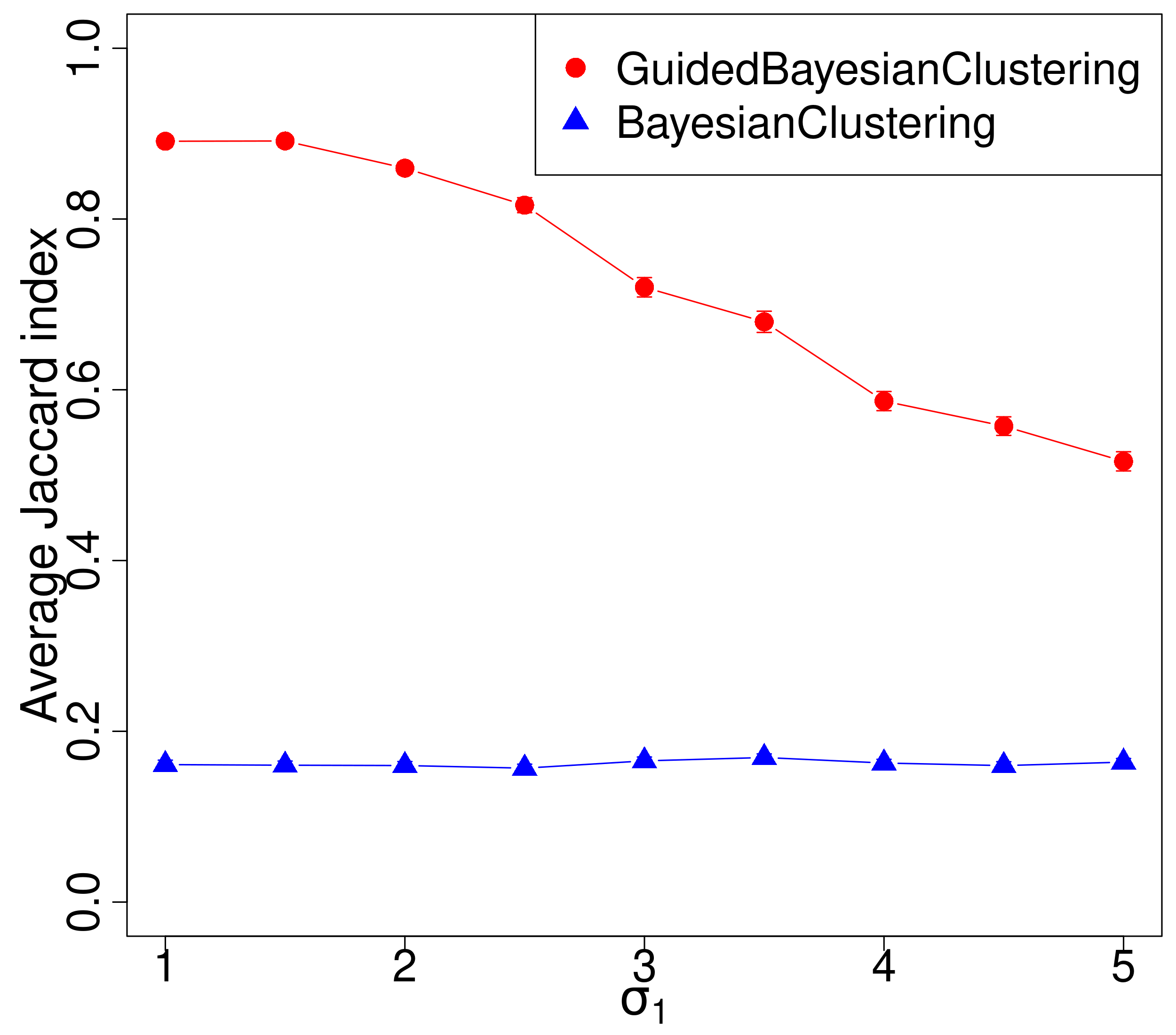}}}

\caption{(a) Using BIC criteria for choosing the number of subtype $K$ based on one simulation with biological variation $\sigma_1=1$; (b-c) the comparison between the GuidedBayesianClustering and the BayesianClustering when varying biological variation $\sigma_1$. Mean estimates of ARI and Jaccard index were presented based on $B=100$ simulations. (b) shows the clustering accuracy assessed using ARI; (c) shows gene selection accuracy assessed using Jaccard index.} 
\label{fig:sim_results}
\end{figure}

For the GuidedBayesianClustering, the clinical guidance term $U_g$ was calculated as the adjusted pseudo R-squared (Section~\ref{sec:VariousOutcome}) from univariate linear regressions.
We follow the description in Section~\ref{sec:prior} to set the hyper parameters unless otherwise specified.  
As shown in Figure~\ref{fig:sim_results}a, the number of clusters was estimated as $K = 3$ using the BIC in Section~\ref{sec:post}.
We controlled the FDR at 0.001 to select genes in both the GuidedBayesianClustering and the BayesianClustering.

Table~\ref{tab:sim} shows the clustering and gene selection results of the two methods with the biological variation $\sigma_1=1$. 
In terms of clustering accuracy, the GuidedBayesianClustering (mean ARI = 0.966) had better performance than the BayesianClustering (mean ARI = 0.161).
Regarding gene selection,
the GuidedBayesianClustering (mean Jaccard index = 0.891) outperformed the BayesianClustering (mean Jaccard index = 0.161).
Since the Jaccard index represented the gene selection accuracy at a specific cutoff (i.e., FDR 0.001) in this scenario,
it was unclear whether the superior performance of the GuidedBayesianClustering was related to this specific cutoff selection instead of the method itself.
We further used the area under the curve (AUC) of a ROC curve to compare the gene selection results of the two methods.
To be specific, we iterated all possible FDR cutoffs and calculated their corresponding sensitivity and specificity for the accuracy of selecting the intrinsic genes.
The AUC of this ROC curve (sensitivity by 1 - specificity) represented the overarching prediction power regardless of a specific FDR cutoff.
As could be expected, the GuidedBayesianClustering (mean AUC = 0.986) outperformed the BayesianClustering (mean AUC = 0.696).
Additionally, Figure~\ref{fig:sim_results}b and Figure~\ref{fig:sim_results}c disclose that the GuidedBayesianClustering still achieved better performance than the BayesianClustering in both clustering and gene selection at different values of biological variation ($\sigma_1=1$ to 5 with an interval of 0.5),
even if the performance of both methods reduced with increasing biological variation.
The superior performance of the GuidedBayesianClustering is expected because it utilized the clinical outcome information to facilitate the identification of clinically relevant subtypes.

\begin{table}[h!]
  \begin{center}
    \caption{Comparison of the performance of GuidedBayesianClustering and BayesianClustering for clustering and gene selection under biological variation $\sigma_1=1$. Mean estimates and standard errors were presented based on $B=100$ simulations. ARI was used to evaluate clustering accuracy. Jaccard index and AUC were used to evaluate gene selection accuracy.}
    \label{tab:sim}
    \setlength{\tabcolsep}{1.8mm}{
    \begin{tabular}{c c c c c}
      \hline
      \hline
           & Clustering results & &\multicolumn{2}{c}{Gene selection results}\\
           & ARI & & Jaccard index & AUC \\
      \hline
      GuidedBayesianClustering  & 0.966 (0.008) & & 0.891 (0.007) & 0.986 (0.001)\\
      BayesianClustering  & 0.161 (0.010) & & 0.161 (0.005) & 0.696 (0.011)\\
      \hline
      \hline
    \end{tabular}}
  \end{center}
\end{table}

\subsubsection{Sensitivity analysis}
\label{sec:sens}

Since we assigned informative prior to $\tau_{\mu0}^2$ and $\tau_{\mu1}^2$ with hyper parameters $a_{\tau_{\mu0}}=2$, $b_{\tau_{\mu0}}=0.005$, $a_{\tau_{\mu1}}=4$, $b_{\tau_{\mu1}}=450$, we conducted sensitivity analysis to check the impacts of the choices of these hyper parameters on the clustering results and gene selection results. 
Each time, we varied one hyper parameter while fixing the other three hyper parameters.
Specifically, we examined 10 evenly-spaced values from 1.1 to 2 for  $a_{\tau_{\mu0}}$; 10 evenly-spaced values from 0.0005 to 0.005 for $b_{\tau_{\mu0}}$; 10 evenly-spaced values from 1.5 to 6 for $a_{\tau_{\mu1}}$; 10 evenly-spaced values from 50 to 500 for $b_{\tau_{\mu1}}$.
As shown in Figure S1, the choices of the four hyper parameters have no effects on the clustering results. 
Figure S2 shows that the choices of the four hyper parameters have little effect on the gene selection results.
Therefore, our algorithm is not sensitive to the perturbation of these informative hyper parameters,
and we will stick with the proposed value of the informative hyper parameters throughout the article.

\subsection{Breast cancer application}
\label{sec:metabric}

In this section, we examined the performance of the GuidedBayesianClustering on real data. 
We applied the GuidedBayesianClustering to METABRIC \citep{curtis2012genomic}, a gene expression dataset for breast cancer containing gene expression profile of 1,870 subjects and 24,368 genes.
For data preprocessing, we filtered out 50\% genes with low average expression levels; scaled the data so that the average expression value of each gene was 0 with a standard deviation of 1; and finally retained 12,180 gene features. 
In this study cohort, various types of clinical outcome variables were measured, including
Estrogen receptor status (ER, binary variable);
HER2 receptor status (HER2, ordinal variable);
Nottingham prognostic index (NPI, continuous variable);
and overall survival.
Each of these four clinical outcome variables served as a guidance for the GuidedBayesianClustering to achieve clustering and gene selection. 
We named the GuidedBayesianClustering with each clinical guidance as ER-GuidedBayesianClustering, HER2-GuidedBayesianClustering, NPI-GuidedBayesianClustering, and Survival-GuidedBayesianClustering, respectively.

To incorporate the information from each clinical outcome variable, we computed the gauidance term $U_g$ as the coefficient of determination $R^2$ or adjusted pseudo-$R^2$, which was derived from the univariate regression model with the expression level of gene $g$ as the response and the clinical outcome variable as the covariate.
Specifically, linear model was applied to the continuous outcome; generalized linear models were used for the binary and ordinal outcome; and the Cox regression model was built on the survival outcome. 
We set the number of subtypes $K=5$ since the PAM50 definition \citep{perou2000molecular} implied there existed 5 subtypes of breast cancer. 
After 3,000 iterations for Gibbs sampling, we discarded the first 1500 iterations as burn-in samples and keep the last 1500 iterations for inference.
To check the convergence of the parameters $\mu_{gk}, \sigma_g, \tau_{\mu0}, \tau_{\mu1}, \tau_{U0}$, and $\tau_{U1}$, we drew trace plots for these parameters using sample after the burn-in period. As shown in Figure S3, (NPI-GuidedBayesianClustering), Figure S4 (ER-GuidedBayesianClustering), Figure S5 (HER2-GuidedBayesianClustering), and Figure S6 (Survival-GuidedBayesianClustering), all these parameters converge to their stationary distribution.  
For a fair comparison between the GuidedBayesianClustering and the non-guided BayesianClustering, we selected number of intrinsic genes to be exactly 400 for both methods (See Table~\ref{tab:evaluate}), which can be achieved by adjusting the FDR criteria. 
This will help eliminate the possibility that the superior performance of a method is due to a greater/fewer number of selected genes compared to the other method.

To benchmark the homogeneity of disease subtype patterns, we utilized Silhouette score \citep{rousseeuw1987silhouettes}, where larger Silhouette score demonstrated not only better separation between clusters but better cohesion within respective clusters as well.
As shown in heatmap patterns, the clustering results from GuidedBayesianClustering (mean Silhouette = 0.051 $\sim$ 0.072, Figure~\ref{fig:heatmapKM}a, ~\ref{fig:heatmapKM}c, ~\ref{fig:heatmapKM}e, ~\ref{fig:heatmapKM}g) were more homogenous than the clustering result from the BayesianClustering (mean Silhouette = 0.033, Figure~\ref{fig:heatmapKM}i)).
Furthermore, since there was no underlying true clustering results, the overall survival difference between subtypes was used to indicate whether the obtained subtypes were clinically meaningful. 
The Kaplan-Meier survival curves for the five subtypes derived from the GuidedBayesianClustering with each clinical guidance were well separated, indicating a significant survival difference ($p=5.74 \times 10^{-12} \sim 1.17 \times 10^{-8}$), Figure~\ref{fig:heatmapKM}b,~\ref{fig:heatmapKM}d,~\ref{fig:heatmapKM}f,~\ref{fig:heatmapKM}h), 
while the BayesianClustering method only achieved moderate significant survival difference ($p=3.48 \times 10^{-6}$, Figure~\ref{fig:heatmapKM}j).

This is expected since all the GuidedBayesianClusering methods had guidance from clinical outcome variables but the BayesianClustering method did not.
Remarkably, the NPI-GuidedBayesianClusering, the ER-GuidedBayesianClusering, and the HER2-GuidedBayesianClusering still achieved good survival separation even though they did not use any survival information. 
This is not unreasonable, as all of these clinical outcome variables were associated with breast cancer and therefore might affect the overall survival.

Since PAM50 is considered the gold standard for breast cancer subtypes, we compared the resulting subtypes obtained by each method with the PAM50 subtypes. 
As shown in Table~\ref{tab:evaluate}, the ARI values from the GuidedBayesianClustering with all four types of clinical outcome guidance (0.223 $\sim$ 0.236) were greater than that from the BayesianClustering (0.176).    
Compared with the BayesianClustering, the subtype results obtained by the GuidedBayesianClustering were more consistent with the gold standard.

\begin{figure}
\centering
  \subfloat[Heatmap from the NPI-GuidedBayesianClustering]{\label{fig:heatmap_NPI}
    \includegraphics[width=.4\textwidth]{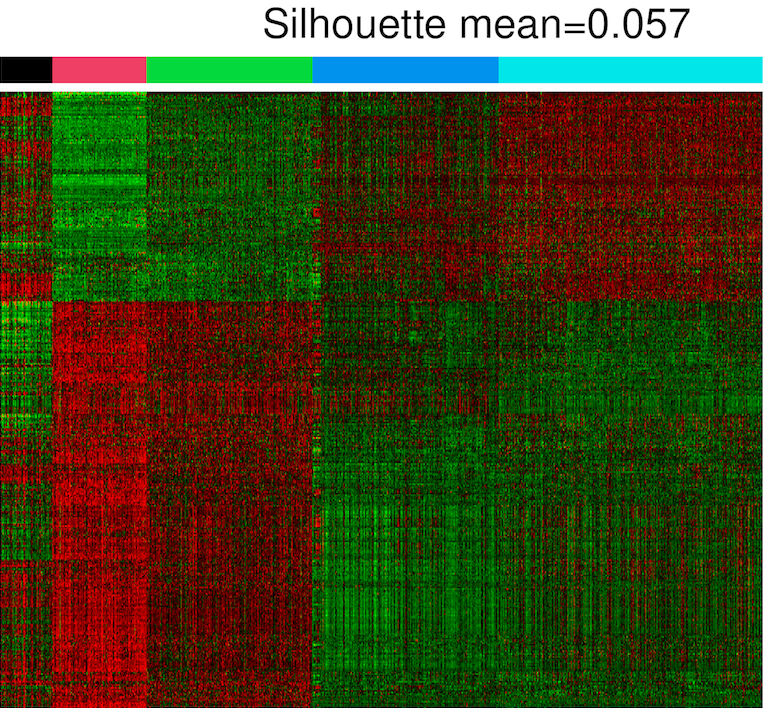}}
~
  \subfloat[Survival curves from NPI-GuidedBayesianClustering]{\label{fig:KM_NPI}
    \includegraphics[width=.4\textwidth]{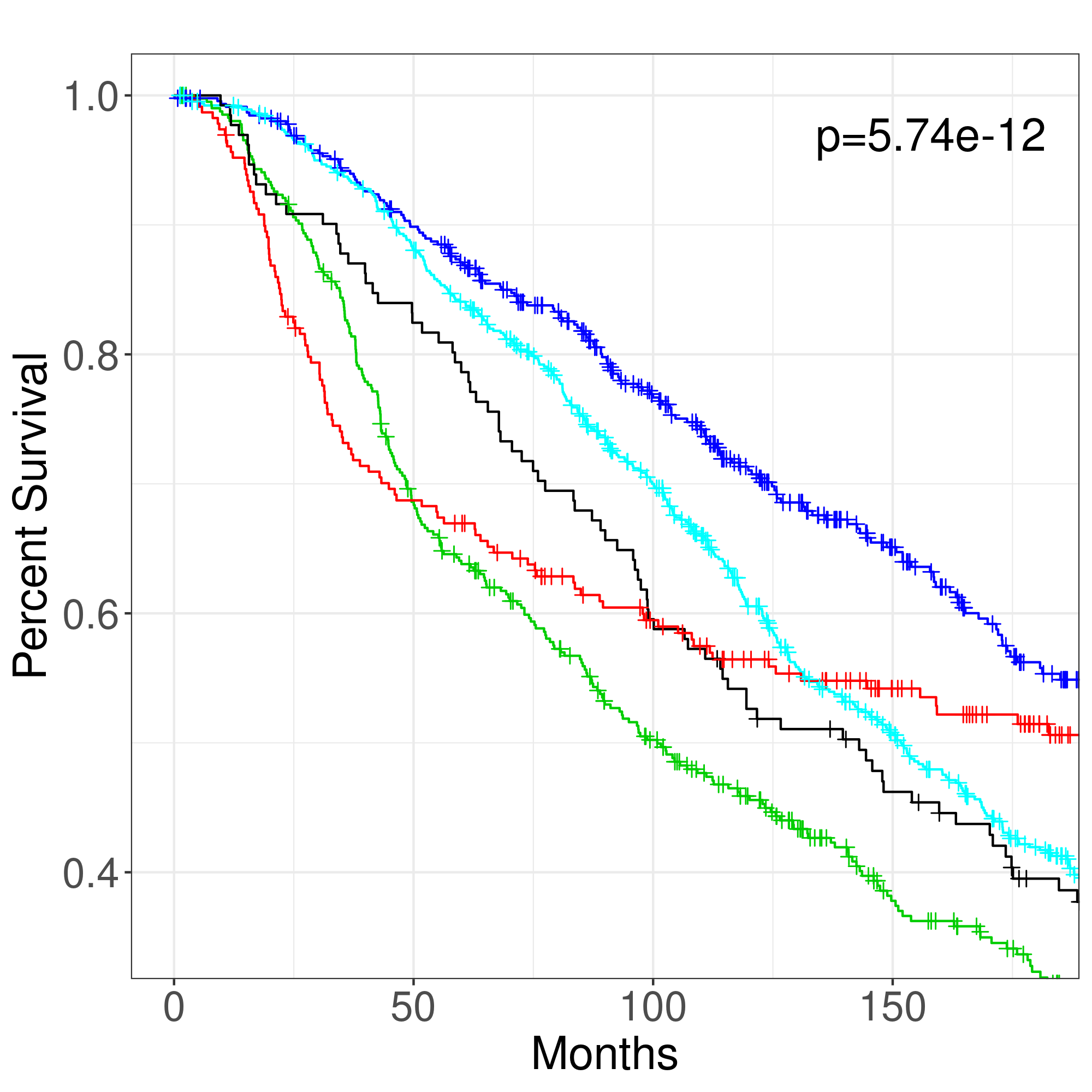}}
    
  \subfloat[Heatmap from the ER-GuidedBayesianClustering]{\label{fig:heatmap_ER}
    \includegraphics[width=.4\textwidth]{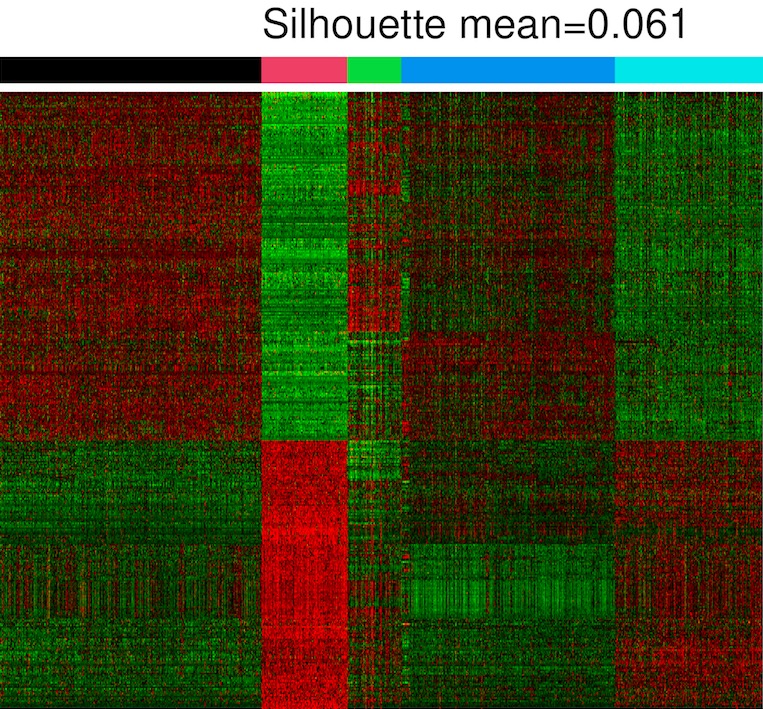}}
~
  \subfloat[Survival curves from ER-GuidedBayesianClustering]{\label{fig:KM_ER}
    \includegraphics[width=.4\textwidth]{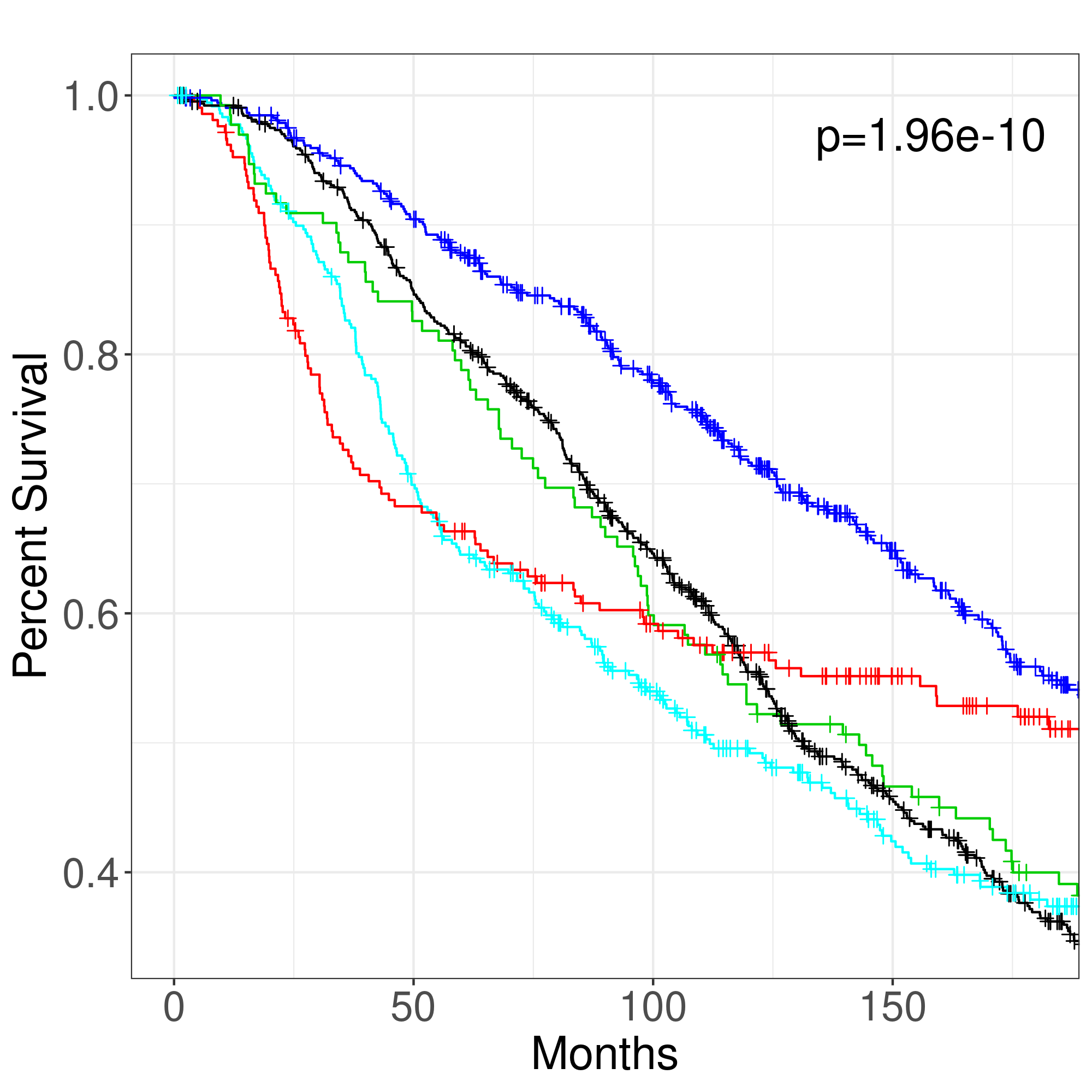}}
        
  \subfloat[Heatmap from the HER2-GuidedBayesianClustering]{\label{fig:heatmap_HER2}
    \includegraphics[width=.4\textwidth]{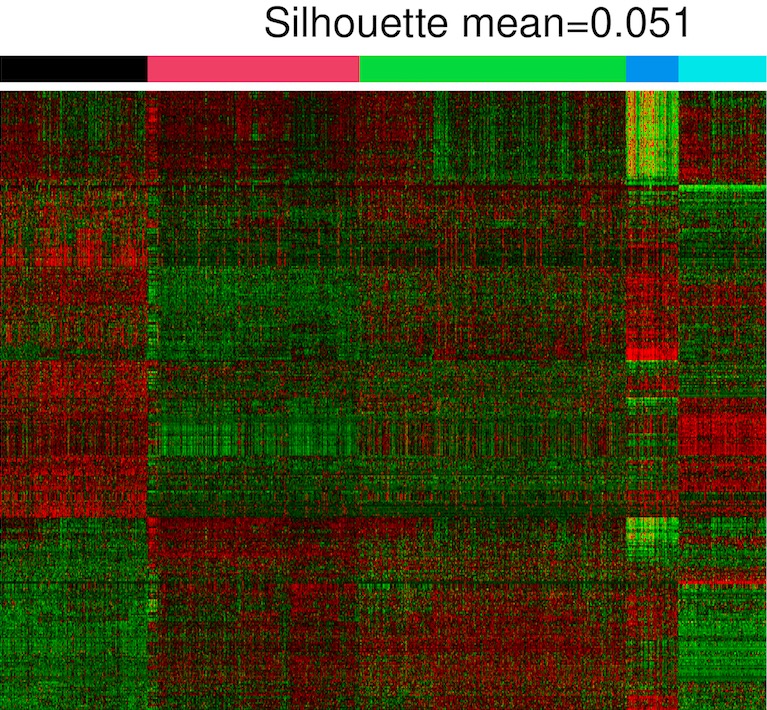}}
~
  \subfloat[Survival curves from HER2-GuidedBayesianClustering]{\label{fig:KM_HER2}
    \includegraphics[width=.4\textwidth]{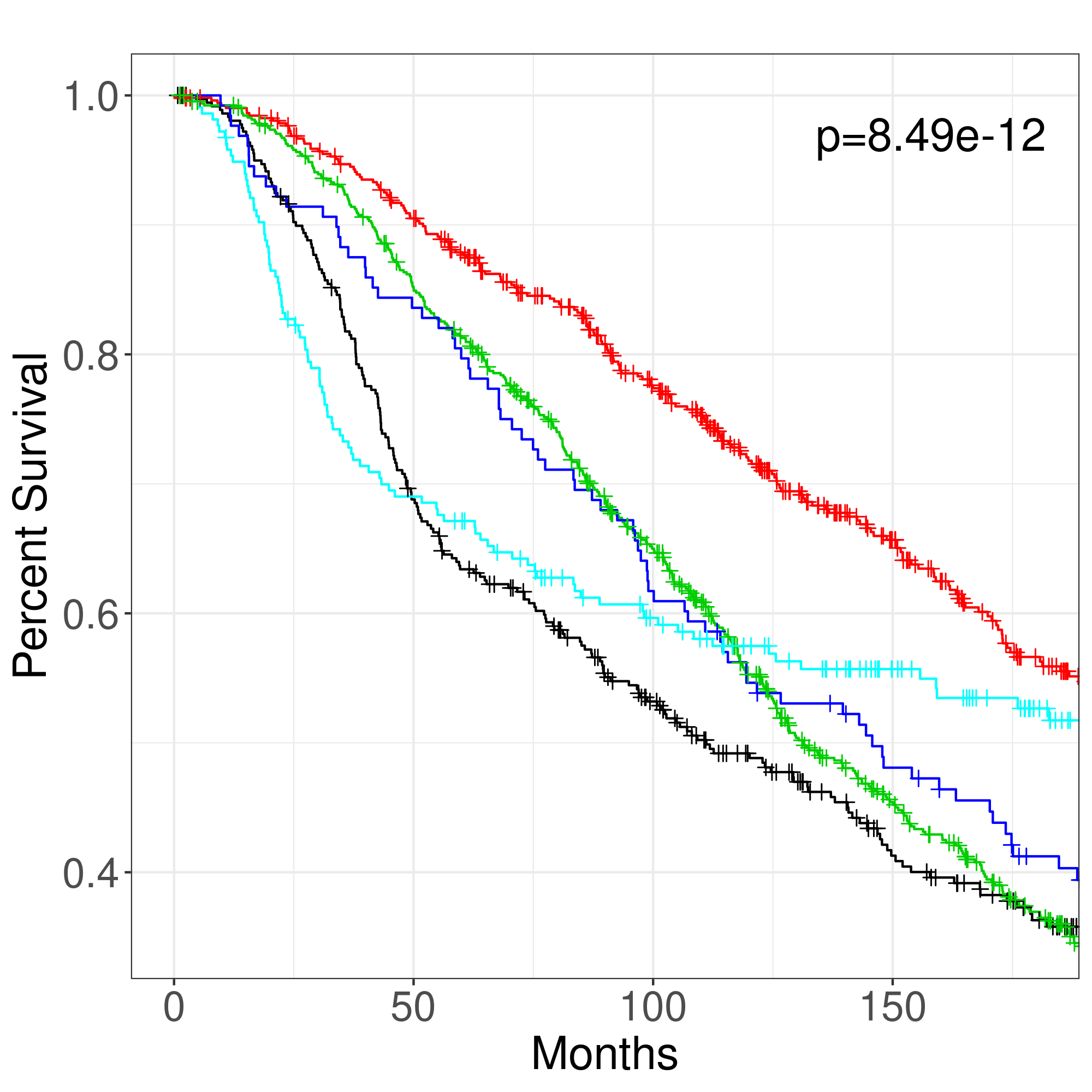}} 
\end{figure}

\begin{figure}
\ContinuedFloat
\centering
  \subfloat[Heatmap from the Survival-GuidedBayesianClustering]{\label{fig:heatmap_Survival}
    \includegraphics[width=.4\textwidth]{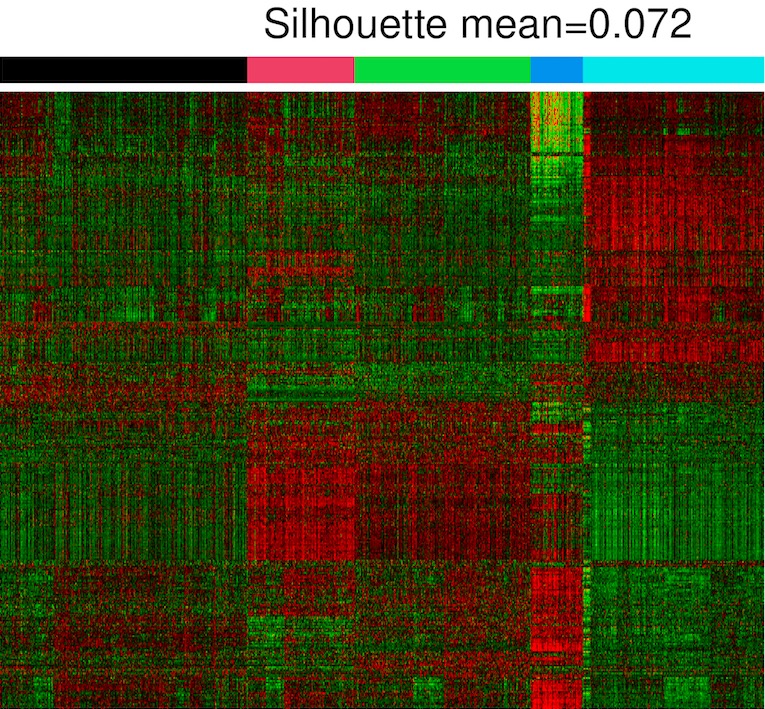}}
~
  \subfloat[Survival curves from Survival-GuidedBayesianClustering]{\label{fig:KM_Survival}
    \includegraphics[width=.4\textwidth]{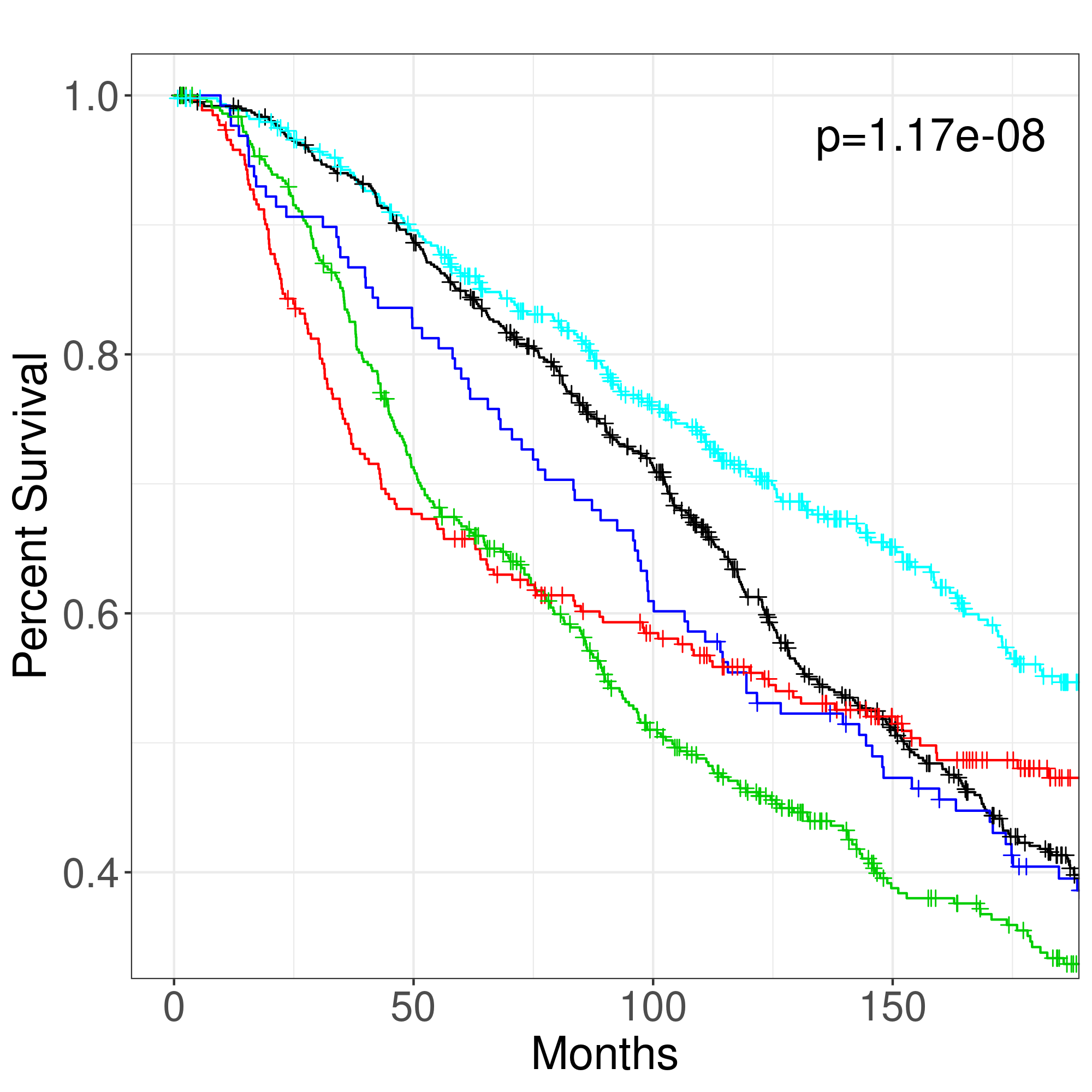}}
    
  \subfloat[Heatmap from BayesianClustering]{\label{fig:heatmap_noGuide}
    \includegraphics[width=.4\textwidth]{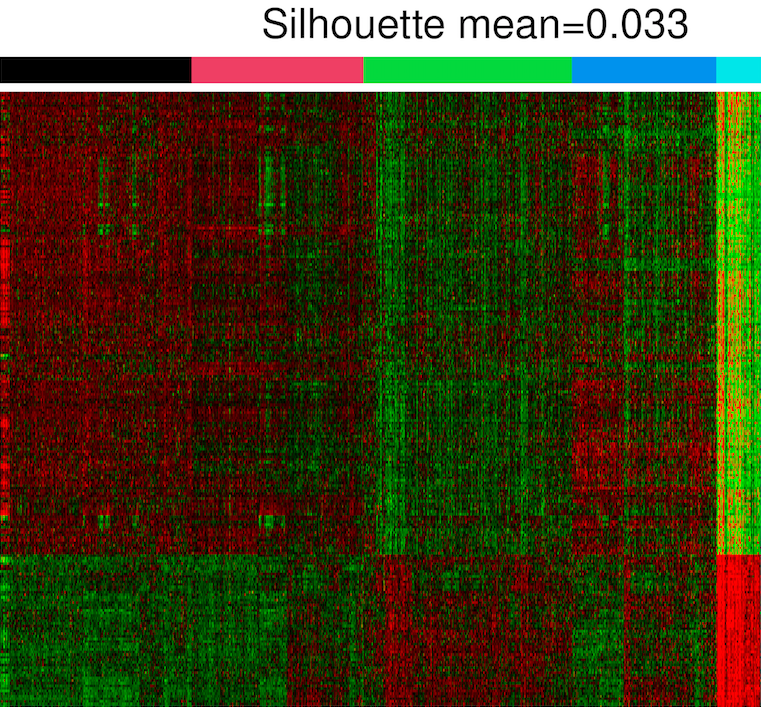}}
~
  \subfloat[Survival curves from BayesianClustering]{\label{fig:KM_noGuide}
    \includegraphics[width=.4\textwidth]{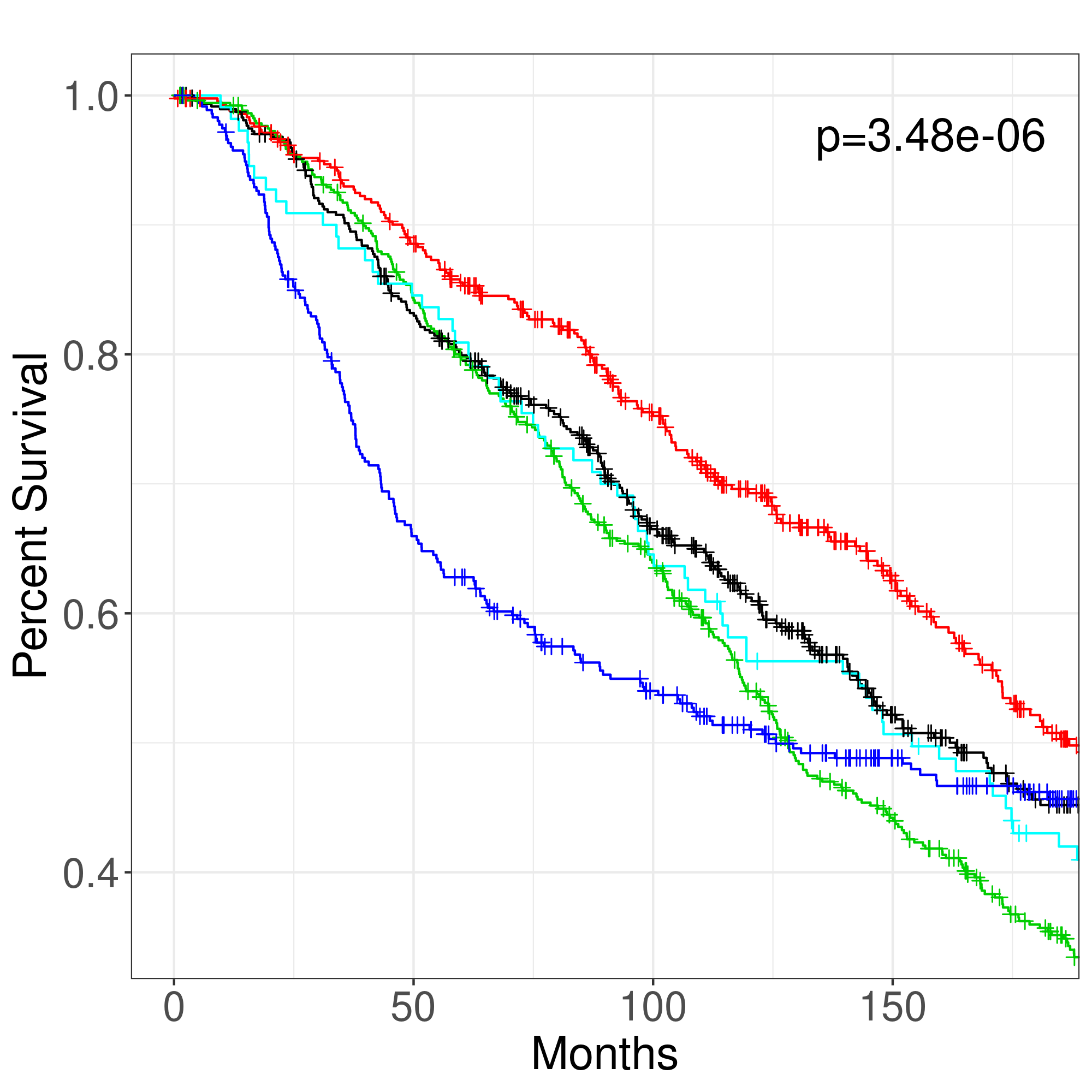}}

\setcounter{figure}{4}

\caption{Gene expression heatmaps with mean Silhouette scores and Kaplan–Meier survival curves with p-values using the NPI-GuidedBayesianClustering, the ER-GuidedBayesianClustering, the HER2-GuidedBayesianClustering, the Survival-GuidedBayesianClustering and the BayesianClustering for the METABRIC dataset. In heatmap (a)(c)(e)(g)(i), rows represent genes and columns represent samples. Red indicates high expression and green indicates low expression. In the bar above the heatmaps, samples are divided into five subgroups with different colors. In survival curves (b)(d)(f)(h)(j), the color of the survival curve for each subgroup corresponds to the color of the subgroup in the heatmap of the same method.}
\label{fig:heatmapKM}
\end{figure}

\begin{table}[h!]
  \begin{center}
    \caption{Comparison of the four GuidedBayesianClustering methods with the non-guided BayesianClustering method in clustering and gene selection. 
    The number of genes selected by each method was 400. ARI was used to compare the clustering results with the PAM50 subtypes. The mean Silhouette score was applied to evaluate the separation and cohesion in respective clusters. P-values for survival differences in the obtained subtypes were calculated using log-rank test.}
    \label{tab:evaluate}
    \setlength{\tabcolsep}{1.5mm}{
    \begin{tabular}{c| c c c c c}
      \hline
      \hline
      Method &Guidance & Genes &ARI &Silhouette &p-value\\
      \hline
              &NPI(continuous) &400 &0.235 &0.057 &$5.74\times10^{-12}$ \\
      Guided  &ER(binary)      &400 &0.234 &0.061 &$1.96\times10^{-10}$ \\
      BayesianClustering &HER2(ordinal) &400 &0.223 &0.051 &$8.49\times10^{-12}$ \\
              &Survival        &400 &0.236 &0.072 &$1.17\times10^{-8}$ \\
      \hline
      BayesianClustering &         &400 &0.176 &0.033 &$3.48\times10^{-6}$ \\
      \hline
      \hline
    \end{tabular}}
  \end{center}
\end{table}

\begin{figure}[h!]
\centering
\includegraphics[scale=0.3]{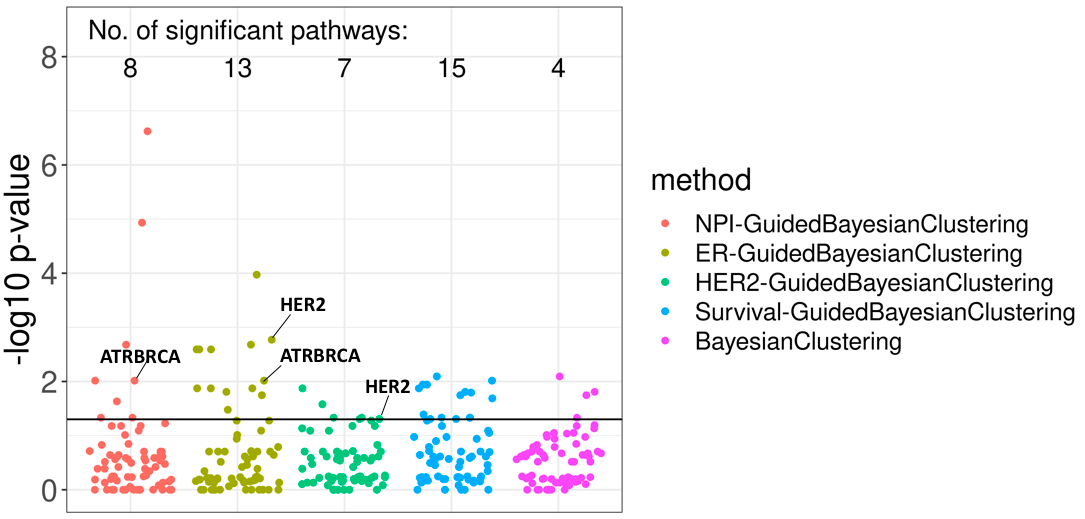}
\caption{Results of pathway enrichment analysis for the GuidedBayesianClustering with four types of clinical guidance, and the BayesianClustering. The number of significant pathways for each method is shown above with the cutoff $p < 0.05$.}
\label{fig:jitter}
\end{figure}

We further performed pathway enrichment analysis to assess whether the selected genes were biologically meaningful, where Fisher's exact test were applied in the BioCarta pathway database. 
The enriched pathways obtained from each method were shown in Figure~\ref{fig:jitter}.
Using $p=0.05$ as cutoff, the number of significant pathways obtained from the NPI-GuidedBayesianClustering (n=8), the ER-GuidedBayesianClustering (n=13), the HER2-GuidedBayesianClustering (n=7) and the Survival-GuidedBayesianClustering (n=15) were more than that obtained from the BayesianClustering (n=4).
This means the genes selected by the GuidedBayesianClustering methods with four clinical outcome variables were more biologically interpretable than those selected by the non-guided BayesianClustering method.
Notably, the ER-GuidedBayesianClustering and the HER2-GuidedBayesianClustering identified HER2 pathway as significant with $p=1.69\times10^{-3}$ and $p=0.049$ respectively, 
which was consistent with previous discoveries of the crucial role of ER and HER2 pathways in relation to  breast cancer \citep{giuliano2013bidirectional}.  
Moreover, the ATRBRCA pathway was significantly associated with genes selected by the NPI-GuidedBayesianClustering ($p=9.65\times10^{-3}$) and the ER-GuidedBayesianClustering ($p=9.65\times10^{-3}$),
Which was also in line with previous studies that the ATRBRCA signaling pathway has strong implications in breast cancer susceptibility \citep{paul2014breast}.
In contrast, none of these hallmark pathways were enriched by the BayesianClustering method.
In conclusion, the GuidedBayesianClustering method has the capability of selecting the most biologically interpretable genes.

\section{Discussion}

We proposed a full a Bayesian framework to identify disease subtypes via high-dimensional transcriptomic data.
Our method is capable of performing gene selection, incorporating the guidance from a clinical outcome variable. 
Various types of clinical outcome variables, including continuous, binary, ordinal and survival data, could be incorporated as guidance in our framework.
Conjugate priors were employed to facilitate the efficient Gibbs sampling.
A decision framework was implemented to infer the false discovery rate of the selected genes.
Through simulations and the application of breast cancer gene expression data, we demonstrated that our proposed method (i.e., the GuidedBayesianClustering method) was superior to the non-guided method (i.e., the BayesianClustering method) in terms of clustering accuracy and gene selection accuracy/interpretability.

One issue with our approach is the proper selection of clinical guidance when multiple clinical outcome variables are available, as only one clinical outcome variable can be accommodated to our method. 
We recommend that the most biologically significant outcome variable be selected as a clinical guide based on available domain knowledge.
On the other hand, if domain knowledge is not available or a data driven approach is preferred, we recommend trying multiple clinical outcome variables separately as guidance. We can then decide to incorporate the outcome guidance which has the most biologically plausible interpretability for the obtained subtypes (i.e., survival difference, pathway analysis). 
In our breast cancer application example, we would suggest to use the results from ER-GuidedBayesianClustering because of (i) good subtype patterns; (ii) significant survival differences between subtypes; and (iii) the largest number of significant enriched pathways.
To further deal with this issue, we will create a multivariate outcome-guided Bayesian clustering method that uses multiple clinical outcome variables as the guidance.

We have implemented our method in the R package ``GuidedBayesianClustering", which can be found on GitHub (https://github.com/LingsongMeng/GuidedBayesianClustering). 
We expect that our method can be applied to identify disease subtypes and select genes for various complex diseases including 
cancer \citep{iyengar2021effects, han2022single}, 
aging and chronic pain \citep{strath2023socioeconomic, peterson2023pain, strath2023vitamin, montesino2022enrichment, cruz2022epigenetic},
gastroenterology \citep{montrose2021dietary}, 
immune disease \citep{drashansky2021bcl11b}, 
and circadian disorder induced disease \citep{wolff2023defining}.   
With the accumulation of high-dimensional omics data and their relevant clinical outcomes, 
our proposed method will be quite applicable to identify clinically meaningful disease subtypes.

\section*{Supplementary Material}

Details for the sensitivity analysis of hyper parameters and the convergence check of parameters are provided in Supplementary Materials for Outcome-guided Bayesian Clustering for Disease Subtype Discovery Using High-dimensional Transcriptomic Data.

\section*{Acknowledgments}
We thank the anonymous reviewers for their valuable suggestions.

\bibliographystyle{tfs}
\bibliography{reference}

\end{document}